\documentclass[final,12pt,sort&compress]{elsarticle}
\usepackage{amssymb}
\usepackage{amsmath}
\usepackage{amsthm}
\usepackage{graphicx}
\usepackage{listings}
\usepackage[all]{xy}
\usepackage{color}
\usepackage{subfigure}
\usepackage{geometry}
\usepackage{color,soul}
\usepackage{multirow}
\usepackage{booktabs}
\usepackage{enumitem}

\begin{document}
	\begin{frontmatter}
	\title{An unstructured block-based adaptive mesh refinement approach for explicit discontinuous Galerkin method}
	\author[ad1,ad2]{Yun-Long Liu}
	\author[ad1,ad2]{A-Man Zhang}
    \author[ad1]{Qi Kong}
    \author[ad1]{Lewen Chen}
	\author[ad1]{Qihang Hao}
    \author[ad1]{Yuan Cao}
	\address[ad1]{College of Shipbuilding Engineering, Harbin Engineering University, Harbin, China}
    \address[ad2]{National Key Laboratory of Ship Structural Safety, Harbin, China}
 
	\begin{abstract}
		In the present paper, we present an adaptive mesh refinement(AMR) approach designed for the discontinuous Galerkin method for conservation laws.
		The block-based AMR is adopted to ensure the local data structure simplicity and the efficiency, while the unstructured topology of the initial blocks is supported by the forest concept such that the complex geometry of the computational domain can be easily treated. 
		The inter-block communication through guardcells are introduced to avoid the direct treatment between flux computing between cells at different refinement levels. 
		The sharp corners and creases generated during directly refinement can be avoided by projecting the boundary nodes to either user-defined boundary surface function or the auto-generated NURBs. 
		High-level MPI parallelization is implemented with dynamic load balancing through a space curve filling procedure.
		Some test cases are presented. As a result, ideal accruacy order and versatility in tracing and controlling the dynamic refinement are observed. Also, good parallelization efficiency is demonstrated.
	\end{abstract}
%
	\end{frontmatter}

	\section{Introduction}

	High-order numerical methods (usually those with an accuracy order of no less than 3) for partial differential equations(PDEs) are attracting more attention from researchers because of their promising advantages. They converge faster, thereby reducing computational costs, especially in long-duration explicit simulations. As one of the most successful high-order schemes, the discontinuous Galerkin (DG) method exhibits good mathematical properties\cite{DG-3,DG-4,JCP_DG_AMR,IJNMF_Tri_VS_quad_DG}. Since it originates from the finite-element method, it can naturally handle complex geometries. Because the basis is not required to be continuous at the cell boundaries, the connection between adjacent cells through the mass matrix is avoided. And,they only connected with their immediate neighbors with the boundary flux terms, which makes the DG method very compact for implementing adaptivity and parallelization. 
	Currently, the DG method has been deeply investigated and widely applied in various fields, esspecially in the computational fluid dynamics, such as the supersonic flow\cite{POF_sup_sonic_flight}, compressible multi-phase flow\cite{JCP_DG_AMR,JCP_WGFM,JSC_2nd_MGFM,JCP_DG_WCW_multiphase,JCP_TGLiu_2phase}, compressible and incompressible viscous flow\cite{JCP_DG_DNS_LES,SIAM_LDG} and fluid structure interactions\cite{OE_DG_FSI,IJNME_DG_FSI}.
	However, it is generally believed to be more computationally expensive in both algebra calculation and RAM consumption compared with other commonly used methods\cite{JCP_DG_AMR}. 
	
	The adaptive resolution technique is an effective approach to scale down the computational cost of the numerical methods for PDEs\cite{JCP_DG_AMR,CPC_paramesh,CPC_p_adaptivity,Kim_ANM_1996}.
	The philosophy is to dynamically adjust the local resolution, either by changing the spatial accracy order or the local mesh size, during the simulation based on proper criterias. The former approach is often referred to as the h-adaptivity while the later one is the so-called p-adaptivity. Even the DG methods are both friendly to the p- and h- adaptivities, in this paper, we mainly focus on the h-adaptivity approach and will refer to as the AMR (adaptive mesh refinement) technique.	 
	The regions that needs to be refined are typically restrains around some moving surfaces in 3-dimensional problems such as the shock wave front and the material interface. These regions are very locallized such that the AMR technique can obtain fine results with much less computational costs than the static grid.

\begin{figure}
		\centering
		\includegraphics[width=\linewidth]{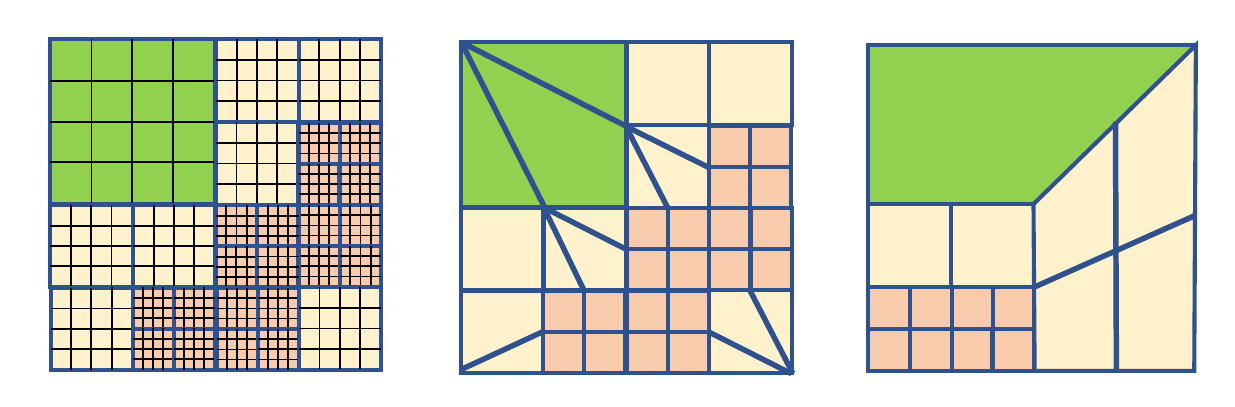}
		\caption{Sketch for different types of AMR. (left) block-structured AMR; (middle) unstructured AMR; (right) tree AMR.}
		\label{fig:amr_type}
	\end{figure}

	Various works have been implemented in the AMR technique, and a lot of packages are available to combine with user-define solvers, e.g., the PARAMESH\cite{CPC_paramesh}, p4est\cite{SIAM_p4est},clawpack\cite{Clawpack}, ENZO\cite{ENZO}, forestDG\cite{JCP_DG_AMR} and so on. As shown in figure \ref{fig:amr_type}, they can be simply classified into 3 types: the block-structured AMR(SAMR) \cite{CPC_paramesh,JCP_patch_AMR}, the unstructured AMR(UAMR) \cite{JCP_UAMR,EWC_ParFum, Basile_CompFluid_2022} and the tree AMR(TAMR) \cite{SIAM_p4est, JCP_DG_AMR}. The SAMR introduces the idea of block which is the basic element of the AMR procedure. Different blocks share the same data structure and structured mesh topology. Hangning nodes exist at the interface between different refinement levels. The blocks at the same refinement level are usually also treated as structured grids at the expense of losing the flexility to deal with complex geometry configurations. On the other hand, the UAMR make proper cell split after refinement to eliminate the hanging nodes which may bring troubles in some discritization methods. But one should note that the hanging nodes are no longer problem for the DG method. The TAMR usually starts from an unstructured mesh and refine each cell without extra cell split to eliminate the hanging nodes. Both of the UAMR and the TAMR trade off their coding simplicity and some performance for flexibility in dealing with different geometries. 
	
	In this paper, we combine the ideas of the SAMR and the TAMR to present a new AMR approach for the explicit discontinuous Galerkin method to solve the conservation laws. Compared with the existing AMR approaches, the present one has the following features:
	 \begin{itemize}
	 \item Block-based AMR with structured internal DG grid to simplify the data structure and gain better performance;
	 \item Supporting unstructured mesh for initial blocks to fit complex geometry boundaries with the forest concept;
	 \item Arbitrary refinement levels to avoid RAM consumption assessment before the simulation;
	 \item High-efficiency Message Passing Interface (MPI) parallelization on HPC.
	 \end{itemize}
	 When a block only contains a single DG cell, the present approach degrades to the TAMR.

	 The rest of the paper is organized as follows. In section \ref{sect:DG-equations}, the basic equations and the idea of the discontinuous Galerkin method are briefly reviewed. Then, the global structure of the present AMR approach is explained in section \ref{sect:framework} to provide a global perspective for the readers. On this basis, we provide a brief introduction about the implementation of the high-level parallelization in section \ref{sect:MPI}. Then, some test cases are given in section \ref{sect:cases}. At last, some conclusions are drawn in section \ref{sect:conclusion}.

	\section{Discontinuous Galerkin method for conservation law system}\label{sect:DG-equations}
	The conserving law is of great significance in many applications. 
	Consider the following conservation-law system in a 2- or 3-dimensional space
	\begin{equation}\label{eq-convserv-equation}
		\frac{\partial \mathbf{U}}{\partial t} + 
		\frac{\partial \mathbf{F}_i}{\partial x_i} 
		 = \mathbf{G}
	\end{equation}
	where $\mathbf{U}$ and $\mathbf{G}(\mathbf{U})$ are the conserving vector and the source vector; $\mathbf{F}_i(\mathbf{U})$ is the flux vector in the $i$th direction. 
	When these vectors are properly chosen, the above system can cover a large range of partial differential equations.
	For example, the well-known compressible Euler equation in 2-dimensional space reads
	\begin{equation}\label{eq-2dEuler}
		\mathbf{U}=\begin{bmatrix}
			\rho \\ \rho u \\ \rho v \\ E
		\end{bmatrix},
		\mathbf{F}_1 = \begin{bmatrix}
			\rho u  \\ \rho u^2 +p  \\ \rho uv  \\ u(E+p) 
		\end{bmatrix},
		\mathbf{F}_2 = \begin{bmatrix}
			\rho v \\ \rho uv  \\ \rho v^2 +p \\ v(E+p)
		\end{bmatrix}
		\text{ and }
		\mathbf{G} =0,
	\end{equation}
	where $\rho$ and $p$ are the fluid density and pressure, respectively. $u$ and $v$ are the velocity components in $x$ and $y$ direction. $E = \frac12 \rho (u^2+v^2) + \rho e$ is the total specific energy with $e$ representing the specific internal energy. The above system can be closed by introducing a proper equation of state to relate $p$ with the conserving vector.

	Next, we will briefly review the basic idea of the Runge-Kutta discontinuous Galerkin method\cite{DG-3,DG-4}. For simplicity, we consider the unknown as a scalar $U$  instead of a vector. 
	Thus, for a single cell $\Omega$ bounded by $\partial \Omega$, the Galerkin form of the control equation reads
	\begin{equation}
		\int_\Omega \frac{\partial U}{\partial t} \phi_i dV  =  
		\int_\Omega (\mathbf{F}  \cdot \nabla \phi_i + 
		S\phi_i) dV - \int_{\partial\Omega} \hat{F} \phi_i dS,
	\end{equation}
	where $\phi_i$ is one of a complete set of the basis functions in the polynomial space $P^K$, $\mathbf{n} $ is the unit normal vector of the cell boundary $\partial \Omega$ pointing outward, and $\hat{F}$ is the numerical flux. 
	By introducing the numerical solution $U_h = \sum_j k_j\phi_j$ to replace $U$ in the above equation, we may rewrite the above equation in the following compact matrix form
	\begin{equation}
		\mathbf{M} \dot{\mathbf{K}} = \mathbf{R},
	\end{equation}
	where $\mathbf{M}$ is the mass matrix with $M_{ij} = \int_\Omega \phi_i \phi_j dV$ , $\dot{\mathbf{K}}$ is the time derivative of the coefficient vector of $k$, and $\mathbf{R}$ is the right-hand side vector which can be explicitly calculated based on the current solution in a time increment.

	Note that the numerical flux should be a single-valued function depending on the solution from both sides of the interface and must satisfy the conditions of consistency, continuity, and monotonicity. A well-known and simple choice is the Lax-Friedrichs flux given by
	\begin{equation}
		\hat{F}(U_L,U_R) = \frac12 \left[\left(\mathbf{F}(U_L)+\mathbf{F}(U_R)\right)\cdot \mathbf{n}-\alpha(U_L-U_R)\right]
	\end{equation}
	where $\alpha$ is the maximum absolute value of the eigenvalues of the Jacobian matrix. Without specifically clarification, the Lax-Friedrichs flux will be used in the tests presented in this paper.
	As for the temperial discritization, we adopt the strong stability-preserving method \cite{SIAM_SSP_time_discretization} to march the solution in time. The details will not be repeated in this paper.

	\section{Basic framework}\label{sect:framework}
	\subsection{Global data structure}
	The present framework is based on the Block-based AMR which has been widely adopted in various packages\cite{CPC_paramesh,AMR_survey,Clawpack,ENZO,AMRex}.
    A block is the basic unit for the adaptive refinement, and each block is divided into cells for the DG simulation. The computational domain is covered by a set of nonoverlapping blocks. They shares the same data structure such that one may only need to implement the DG solver for the cells inside a single block, and the framework loops over all blocks and takes care of the communication between them. 
	The main difference from the other packages with block-based AMR is that the initial root blocks can be connected with an unstructured topology so that complex geometrical computational domains can be easily incorporated.  

\begin{figure}
		\centering
		\includegraphics[width=\linewidth]{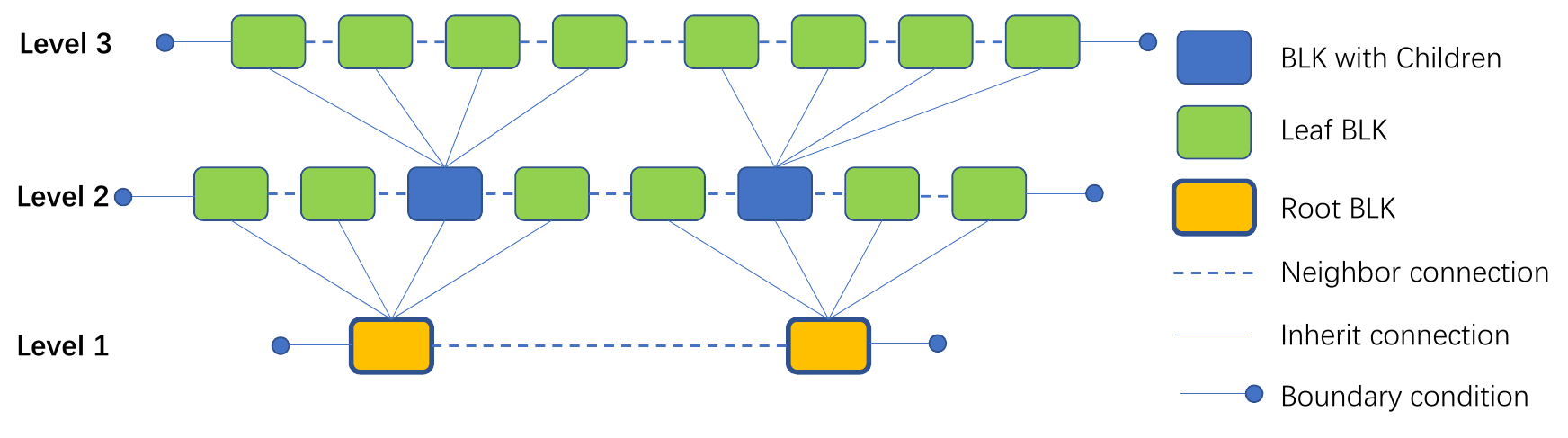}
		\caption{Sketch for the relations inside a block forest taking 2-dimensional problem as an example.}\label{fig:quadtree}
    \end{figure}

    The block is defined as a derived type of the modern Fortran, which contains some scalar and logical indicators, data arrays holding the intermediate variables and the unknowns, and pointers linking with other related blocks, e.g., its parent, children, and immediate neighbors at the same refinement level. 
	All the blocks descendants from the same initial block constitute a block tree, and all the trees connected with each other forms a forest, as shown in figure \ref{fig:quadtree}. 
    The root blocks are denoted by the yellow squares, which have no parents and are at refinement level 1. They are created at first with user-given mesh data, and their neighboring relationship is calculated according to the topological connection. Then, they may be recursively refined to generate children blocks at the next refinement level if necessary. Thus, a quad- or oct-tree structure grows for each root block holding the data structure. The neighboring relationship of the newborn blocks is calculated with the neighboring relationship of their parents and their location in their cousins.

\begin{figure}
		\centering
		\includegraphics[width=0.5\linewidth]{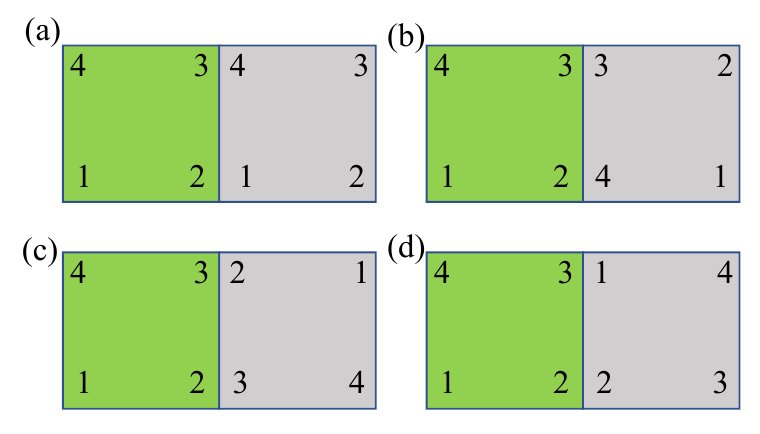}
		\caption{Example sketch for the relative rotation between neighboring blocks. Taking the second neighbor block of the green block as an example, the relative rotations are (a)$R_2 = 0$; (b)$R_2 = 1$; (c) $R_2=3$ and (d) $R_2=4$, respectively. }\label{fig:block-rotation}
    \end{figure}

	Because the root blocks may be defined with the unstructured mesh, the relative rotation between neighboring blocks can be arbitrary. Thus, the array $\mathbf{R}$ is used to record the relative rotation information of a block with its neighbors. $R_i$ indicates the relative rotation number of the $i$th neighbor block with respect to the current one, as shown in figure \ref{fig:block-rotation} for a simple description of a 2-dimensional case. As for the 3-dimensional case, the rotation number ranges from 0 to 23 and follows  the encoding method of \citet{JCP_benson_his}. The rotation number uniquely identifies the topological relationship between the current block and its neighbor, essential to the correct guarding cell filling procedure described in section \ref{sect:guardfill}.

	\subsection{Block geometry conversion from special root blocks}
\begin{figure}
		\centering
		\includegraphics*[width=0.8\linewidth]{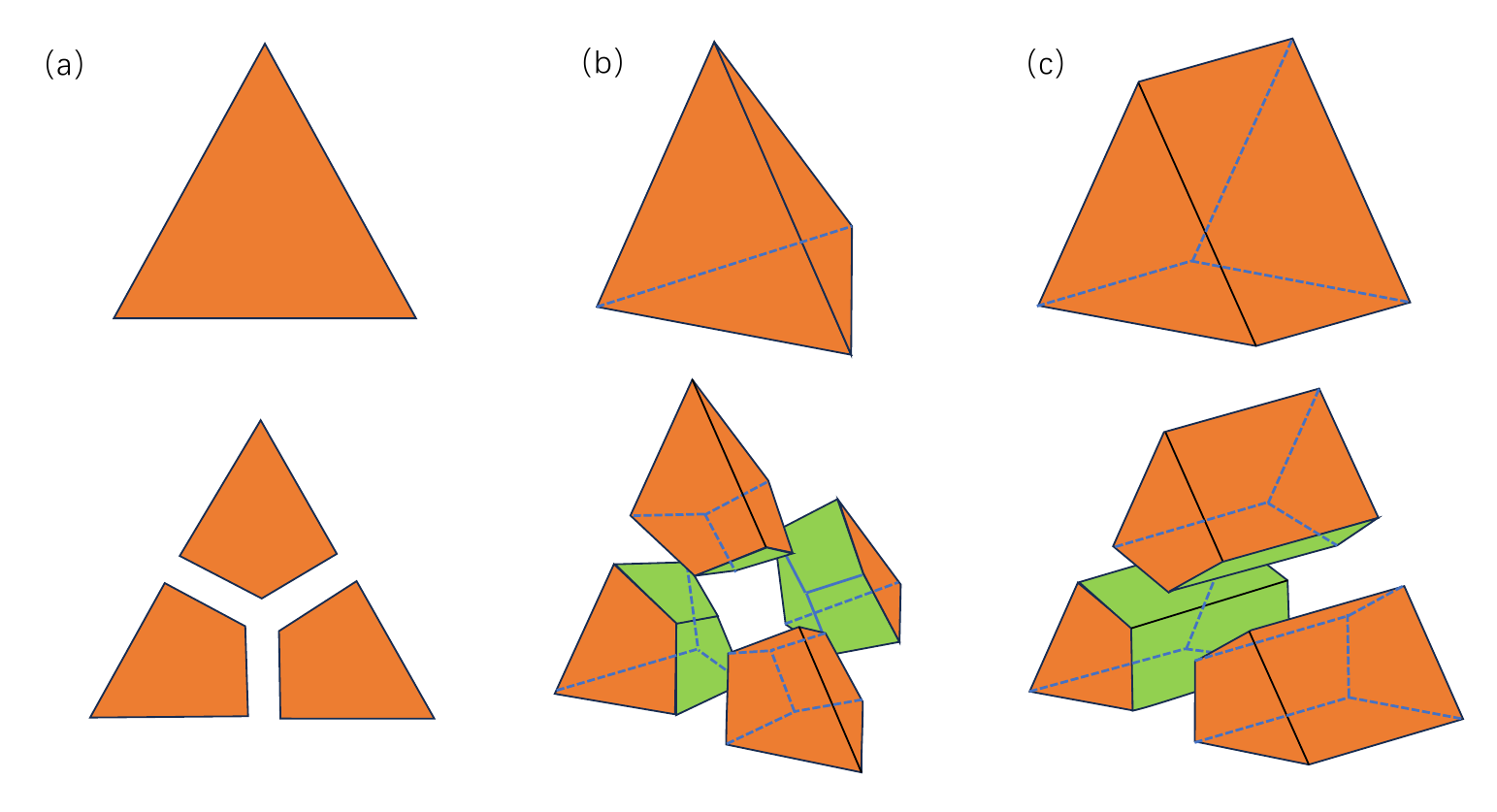}
		\caption{Sketch for the root block type conversion.}
		\label{fig:mesh-convert}
	\end{figure}

	In the present approach, the quadrilateral blocks for 2-dimensional problems and hexahedrons for 3-dimensional problems are directly supported for better performance\cite{IJNMF_Tri_VS_quad_DG} and simplicity. However, it is non-trivial to generate high-quality mesh for problems with complex boundaries. If other types of initial blocks are used to cover the computational domain, we can convert them to quadrilateral or hexahedrons by simply splitting properly, as shown in figure \ref{fig:mesh-convert}. For example, a triangular 2-dimensional block can be split into 3 quadrilateral blocks, while a tetrahedron and a triangular prism can be converted into 4 and 3 hexahedrons, respectively. However, if the initial root blocks are of mixed types, they must be refined at least once to make all the blocks quadrilateral or hexahedrons.

    \subsection{Mesh definition inside a block}
\begin{figure}
		\centering
		\includegraphics[width=0.7\linewidth]{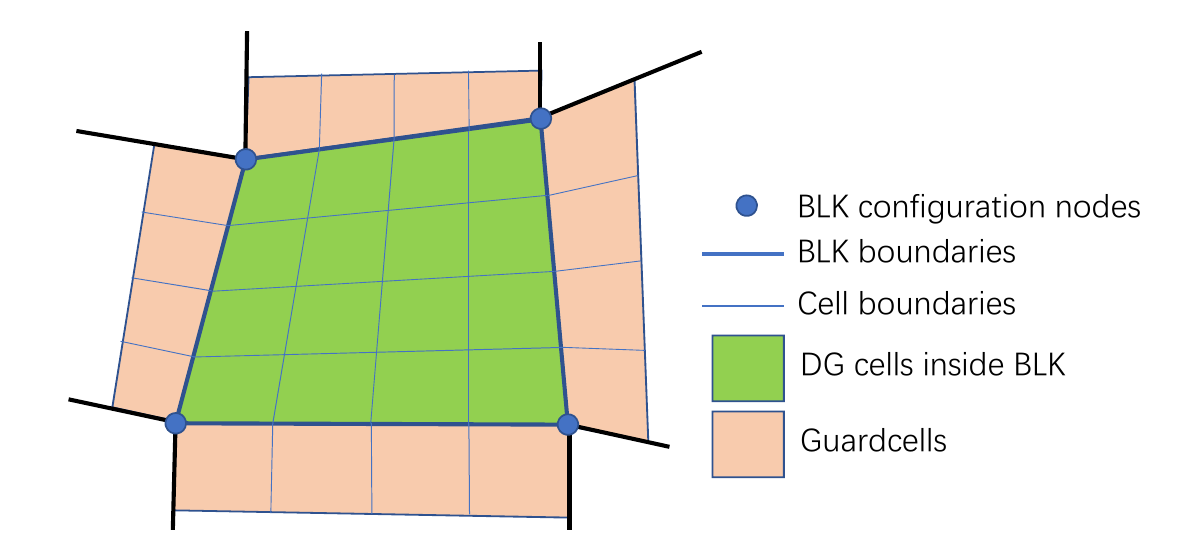}
		\caption{Configuration of mesh definition of a block}\label{fig:block-mesh}
	\end{figure}

     The geometry configuration of each block is defined by its configuration nodes located at the block corners, and is discretized into $N_{seg}^{n_d}$  cells with their nodes interpolated from the block configuration nodes, as shown in figure \ref{fig:block-mesh} for $n_d = 2$ and $N_{seg} = 4$. Here, $n_d$ is the number of the spatial dimensions. For linear geometry configuration, the bilinear (tri-linear for 3-dimensional problem) interpolation is adopted to project any location  $\zeta, \eta$  from the reference space to the physical space, which reads
	 \begin{equation}
		\mathbf{r} = \frac14 (1+\eta_i \eta)(1+\xi_i\xi) \mathbf{r}_i
	 \end{equation}
	 where $\mathbf{r}_i$ is the position vector of the configuration node $i$, $\eta_i$ and $\xi_i$ are its coordinates in the reference space and evenly spaced between -1 to 1 to form a locally structured DG mesh. 
	 
	 As shown in figure \ref{fig:block-mesh}, we also define a layer of guarding cells outside of each block boundary to communicate between neighboring blocks. The node positions of these guarding cells are interpolated from the corresponding neighbor block. 
	 Only the real cells inside the block will be solved during each time increment of solving the DG formulation, before which the solution of guarding cells is directly retrieved from the neighbor blocks to ensure a correct numerical flux at the block boundaries, which will be explained in section \ref{sect:guardfill}.

	 When a block is refined, the children's configuration nodes will be interpolated from the parents with bilinear interpolation. At the same time, the unknowns are calculated with the $L^2$ projection from the parent DG cells, which will be explained in section \ref{sect:datainherit}.
	 
\begin{figure}
		\centering
		\includegraphics*[width=\linewidth]{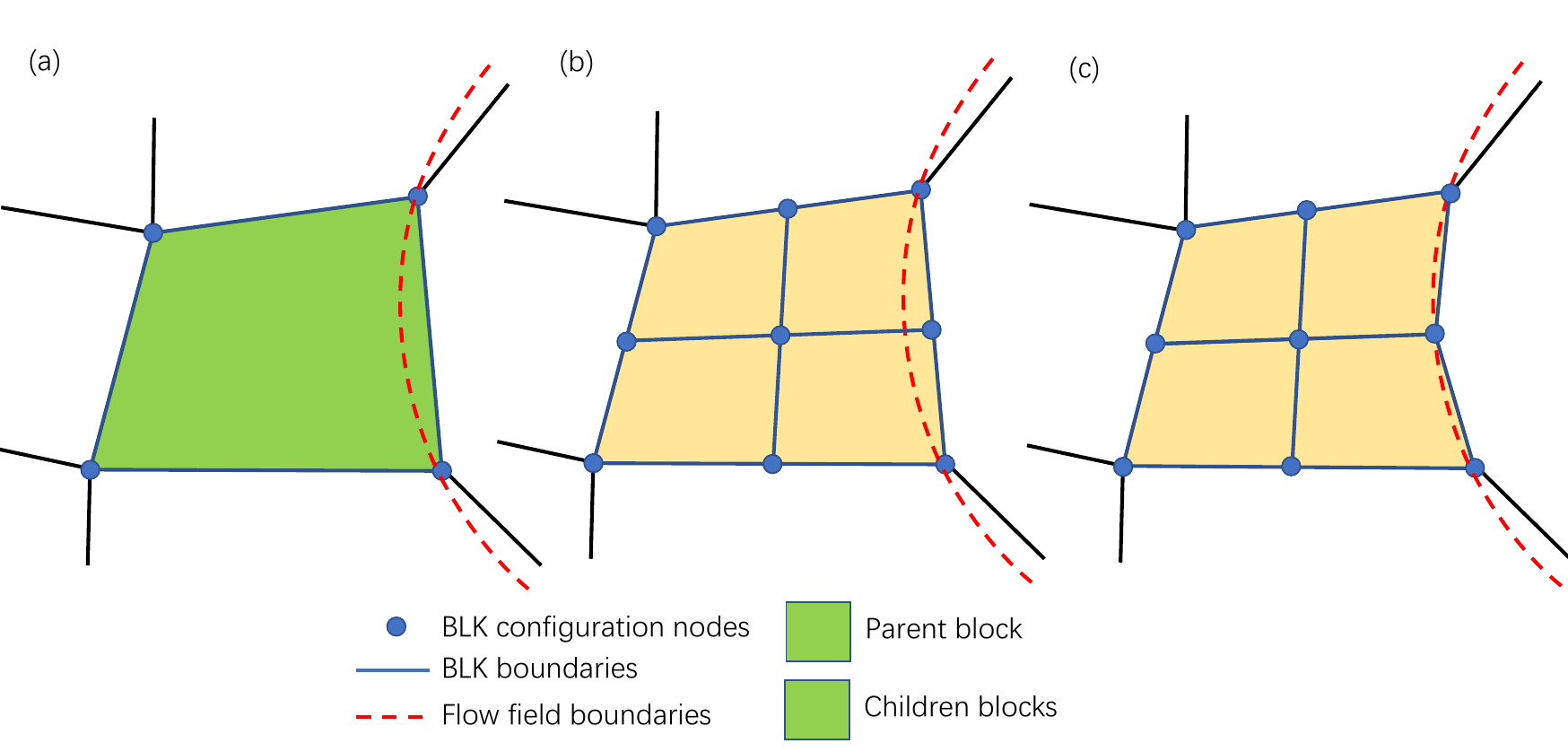}
		\caption{Sketch for the refinement of a given block adjacent to the boundary of the computational domain. (a) the initial block to be refined (b) directly refinement with interpolation (c) boundary configuration nodes adjustment.}
		\label{fig:blkrefine}
	 \end{figure}

	 Note that if a block is adjacent to the exterior boundary of the computational domain which is a curved surface, the bilinear/trilinear interpolation of the cell nodes from the configuration nodes or the configuration nodes of the child block from those of the parent block will lead to sharp edges derivating from the real boundary, as shown in figure \ref{fig:blkrefine}. In such cases, we provide two options to resolve the problem. The first option is to project these newly generated nodes to the computational domain boundary if an analytical equation for the boundary is explicitly given. The projection is implemented with the following Newtonian iterative method:
     \begin{equation}\label{eq-bound-proj}
         \mathbf{r}^{(n+1)} =\mathbf{r}^{(n)} - \frac{\nabla f(\mathbf{r}^{(n)})}{|\nabla f(\mathbf{r}^{(n)})|} f(\mathbf{r}^{(n)})
     \end{equation}
     where $f(\mathbf{r})=0$ is the surface equation for the boundary and $n$ is the number of the iterations. Typically, several times of iterations will produce satisfying results.
	 
\begin{figure}
		\centering
		\includegraphics[width=0.5\linewidth]{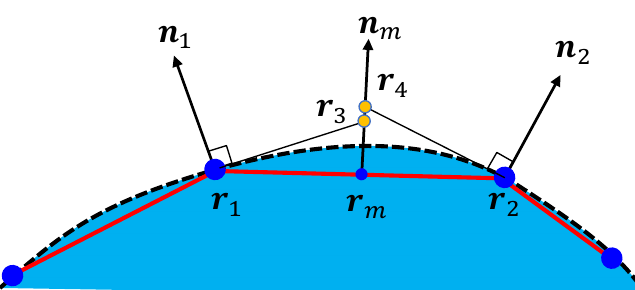}
		\caption{Sketch for the boundary curve reconstruction based on the NURBs in 2-dimensional problem.}
		\label{fig:nurbs2d}
	 \end{figure}

     The second option is used for the computational domain boundary without an analytical surface equation. In such cases, the second-order NURBS is used to calculate the position of the newly generated configuration nodes. In 2-dimensional problems, the parametric equation of the boundary curve is given by
     \begin{equation}
         \mathbf{r} = (1-\theta) [(1-\theta)\mathbf{r}_1 + \theta \mathbf{r}_3] + \theta [(1-\theta)\mathbf{r}_4 + \theta \mathbf{r}_2],
     \end{equation}
	 as shown in figure \ref{fig:nurbs2d}, where $\theta$ is the scaled parameter ranging between 0 and 1, and should be taken as 0.5 here. $\mathbf{r}_1$ and $\mathbf{r}_2$ are the positions of the two configuration nodes of the block boundary, $\mathbf{r}_3$ and $\mathbf{r}_4$ are the control points of the NURBs which are sets to the intersection points of the two tangential lines at $\mathbf{r}_1$ and $\mathbf{r}_2$ and the line pointing from the center point between them with the direction of $\mathbf{n}_m$. Here, $\mathbf{n}_m$ is the algbratic averaged normal vector at $\mathbf{r}_m$.
	 The normal vectors of the configuration nodes is obtained with the following weighted average of the surrounding boundary faces
     \begin{equation}
         \mathbf{n} = \frac{1}{\sum_i 1/S_i} \sum_i \frac{1}{S_i}\mathbf{n}_i
     \end{equation}
     where $\mathbf{n}_i$ and $S_i$ are the normal vector and area of the boundary face $i$ adjacent to the current block configuration node.
	 
     As for the 3-dimensional cases, the surface equation is given in the following form
     \begin{equation}
         \mathbf{r}(\zeta,\eta) = \sum_i  \sum_j N_i(\zeta,\eta)N_j(\zeta,\eta) \mathbf{r}_{ij},
     \end{equation}
     where $N_i$ is the bilinear interpolation weights, and $\mathbf{r}_{ij}$ are the control points for the NURBS surface.

\begin{figure}
		\centering
		\includegraphics[width=0.8\linewidth]{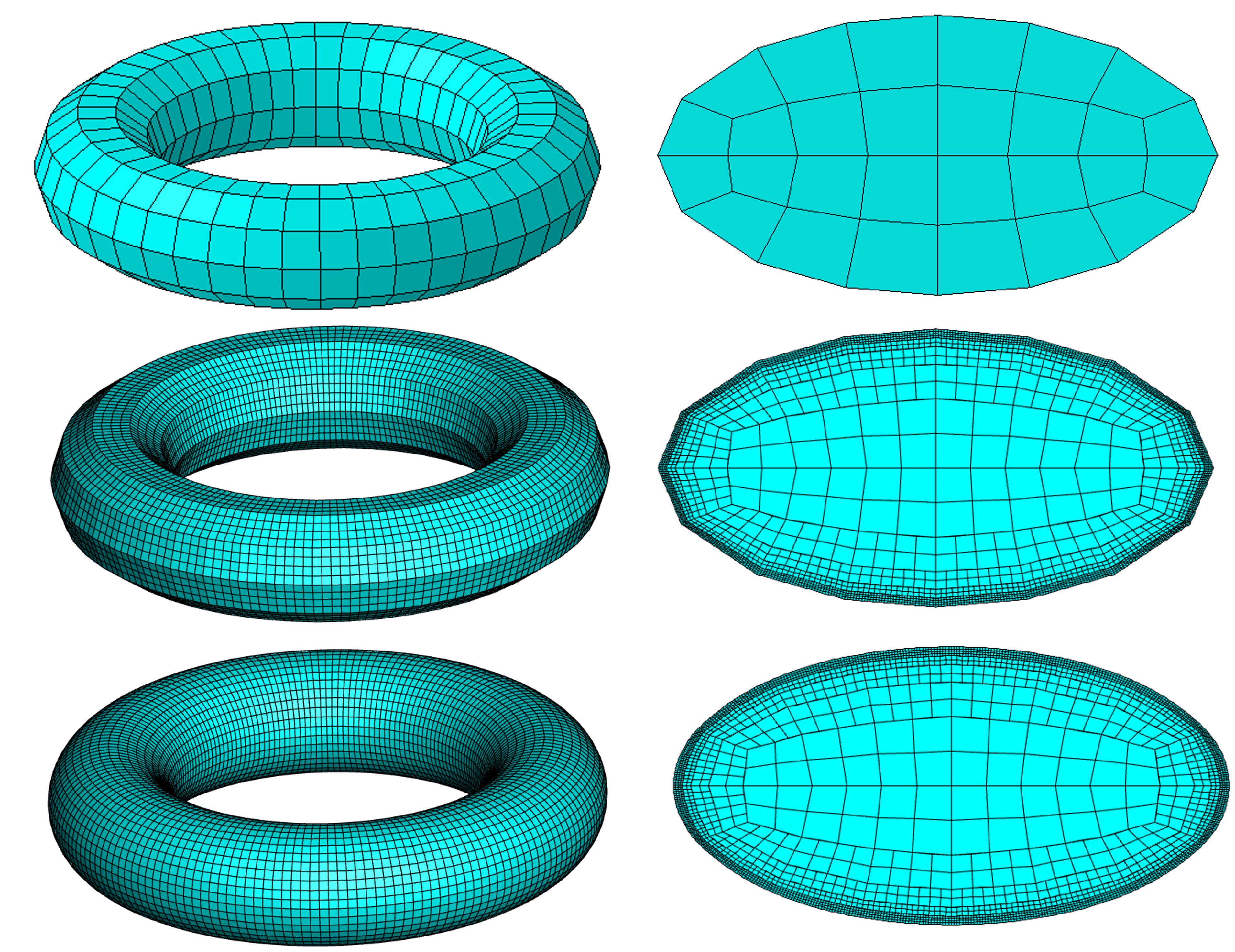}
		\caption{Comparision of the computational boundary of a toroidal and a elliptic computitional domains without (midlle) and with (bottom) the node position projection.}\label{fig:bound-project}
	 \end{figure}

     Two test cases are given in Figure \ref{fig:bound-project} where a toroidal and an elliptic computational domains are refined without and with the boundary projection method. We can see that it avoids the undesired sharp edges on the refined boundaries to approximate the smooth geometry of the computational domain.
	 
	\subsection{Data inherition between imediate relative blocks}\label{sect:datainherit}
	
	When a block is to be refined, $2^{n_d}$  new children blocks are generated. The newborn child block inherites the cell data from its parent with the $L_2$ projection method. For example, the coefficient vector $\mathbf{K}_{C}^{(i)}$ of the $i$th newborn child cell is given by solving the following linear system
	\begin{equation}\label{eq-p2c-inherit}
		\mathbf{M}_{C}^{(i)}\mathbf{K}_{C}^{(i)} = \mathbf{R}^{(i)} 
	\end{equation}
	where $\mathbf{M}_{C}^{(i)}$ is the mass matrix of the child cell $i$. With $\phi^C$ and $\phi^P$ representing the basis of the child and the parent cells, respectively, the right-hand side vector $\mathbf{R}^{(i)}$ can be calculated with 
	\begin{equation}
		R^{(i)}_{\ell} = \int_{\Omega_i} \phi^C_\ell \phi^P_j k_j^P dV.
	\end{equation}

	On the contrary, a parent block will inherit the cell data from all of its children blocks when the a derefined. Thus we have the coefficient vector of the parent cell calculated by solving 
	\begin{equation}\label{eq-c2p-inherit}
		\mathbf{M}_{P} \mathbf{K}_P = \sum_{i=1,N_c}\mathbf{R}^{(i)},
	\end{equation}
	here $\mathbf{R}^{(i)}$ can be calculated with
	\begin{equation}
		R^{(i)}_\ell = \int_{\Omega_i} \phi^P_\ell \phi^C_j k_j^C dV.
	\end{equation}

	\subsection{Communication between neighbor blocks}\label{sect:guardfill}
    Because each block is a single computational domain, communication between neighbor blocks must be implemented to ensure their interaction. The communication is done by filling the guarding cells with the data extracted from the neighbor block. One should note that the guarding cell filling procedure is performed right before the corresponding block is solved. Then data inconsistency may occur, because some source data for the filling procedure may already be updated if the neighbor block is solved before the current one. Thus, following the treatment in PARAMESH\cite{CPC_paramesh}, a temporary array is used to store the un-updated data at the beginning of a time increment and remains unchanged until all the blocks are updated. In such a way, the data consistency can be ensured.

	During the guardcell filling procedure, the relative rotation bewteen neighbor blocks should be treated. When filling data from a source cell with relative rotation, the source cell and the destination cell will have different reference coordinate system. Then, a $L_2$ projection will be used to consider the rotation. 
	Denoting $\mathcal{R}^n$ as the rotating operator to rotate the cell geometry configuration by $n$. Thus, we need to apply the rotation during the guard cell filling. There are 3 types of guard cell filling procedures shown in figure \ref{fig:commu-neigh}:

	\begin{figure}
		\centering
		\includegraphics[width=0.3\linewidth]{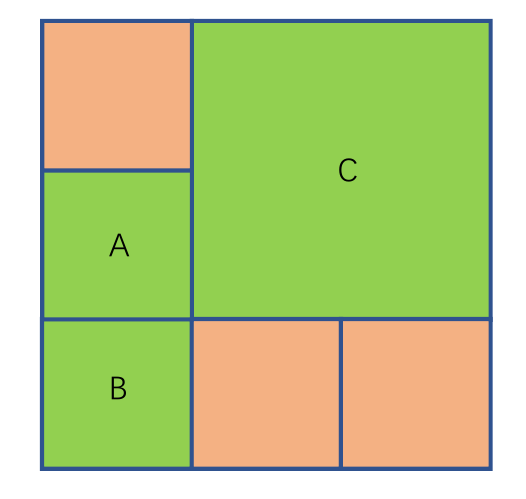}
		\caption{Sketch for the communication between neighboring blocks.}\label{fig:commu-neigh}
	\end{figure}

	\begin{enumerate}[label = (\arabic*)]
		\item Filling with a neighbor block at the same refinement level (From block A to block B in figure \ref{fig:commu-neigh} as an example);
		
		In this case, the solution vector $\mathbf{K}$  of the destination guarding cell $d$  is directly copied from the corresponding cell $s$ of the neighbor block with appropriate rotation manipulation:
		\begin{equation}
			\mathbf{K}_d = \mathcal{R}^n\mathbf{K}_s
		\end{equation}
		where the superscript $n$ is the relative rotation between the neighbor block and the current block.

		\item Filling with a coarser neighbor block(From block C to block A in figure \ref{fig:commu-neigh} as an example).
		\item 
		In this case, the solution of the destination guard cell is obtained by splitting the corresponding parent in the neighbor block:
		\begin{equation}
			\mathbf{K}_d = \mathcal{R}^n \mathcal{S}^i \mathbf{K}_s,
		\end{equation}
		where $\mathcal{S}^i$ is the split operator to calculate the solution data of the $i$th child cell from its parent cell and one may refer to equation (\ref{eq-p2c-inherit}) for detail.

		\item Filling with a finner neighbor block(From block A to block C in figure \ref{fig:commu-neigh} as an example);
		
		In this case, the solution of the destination guard cell is obtained by combining the $N_C$  children cells in the neighbor block:
		\begin{equation}
			\mathbf{K}_d = \mathcal{R}^n \mathcal{C}(\mathbf{K}_s^{(1)},\mathbf{K}_s^{(2)}...\mathbf{K}_s^{(N_C)}),
		\end{equation}
		where $\mathcal{C}$ is the combination operator to combine the $N_C$ children data into the parent cell and one may refer to equation (\ref{eq-c2p-inherit}) for detail.
	\end{enumerate}

	\section{Implementation of parallel computing}\label{sect:MPI}
	\subsection{Dynamic load-balance}
	the load balancing procedure is crucial to large-scale paralleling computation. In the present paper, we provide two approaches to balance the computational loads between CPUs. 

	The first option is the partitioning schemes implemented by the METIS package which is commonly used in finite element meshes. It relies on a random coarsing procedure such that it will give new partitioning results that is unrelated to previous ones even if only a small portion of the computational mesh changes, which is commonly observed in the dynamic AMR procedure. As a result, the communication cost would extensively increase. Thus, this option is recommended only when the AMR procedure is completed before the simulation and the block topology remains static thereafter.

\begin{figure}
		\centering
		\includegraphics[width=0.7\linewidth]{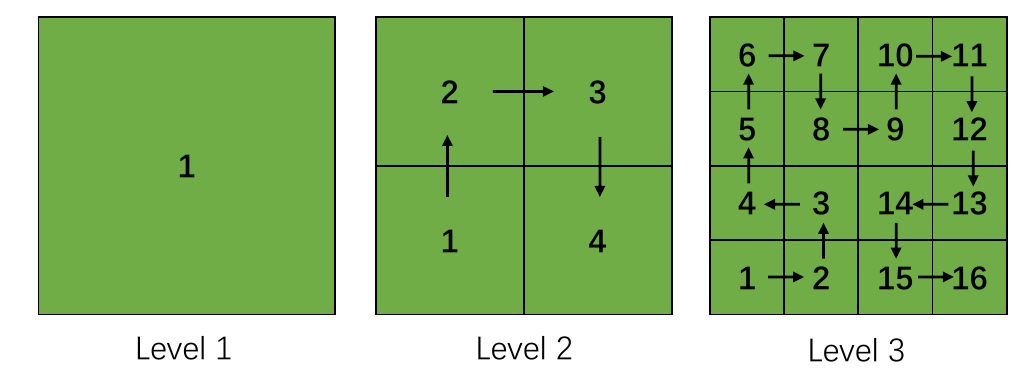}
		\caption{Sketch for the recursive refining of a Hilbert curve in a 2-dimensional domain for the first 3 levels.}\label{fig:hilbert-refine}
	\end{figure}

	As long as the AMR is required dynamically during the simulation, the second option for the load balance, i.e., the SCF(space curve filling) technique, will be prefered. In this package, we implemented the Morton curve and the Hilbert curve for the space filling. They are both self-similar and recursive to map between a coordinate sequence a  higher dimensional space to a one-dimensional space. Thus, we divide the sequence in the one-dimensional space linearly to ensure the coordinates in the same partition in the high-dimensional space are close to each other. An example of the Hilbert curve to fill a 2-dimensional computitional domain is shown in figure \ref{fig:hilbert-refine}. The basic idea is the build a square root block holding all the points initially and recursively do the refinement until each leaf block has at most 1 point inside. Then, the Hilbert space curve passing all the leaf blocks is used to order the points inside. By partitioning the curve with proper weight on each point, the target blocks are partitioned with the fact that the blocks in the same partition are adjacent to each other.
	The SCF technique can incrementally change the partition with minimum data communication requirements when the mesh is locally refined or derefined. 

\begin{figure}
		\centering
		\includegraphics[width=0.5\linewidth]{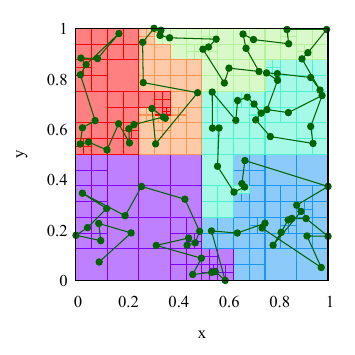}
		\caption{Hilbert space curve filling to reorder arbitrary nodes in 2D space. The color of the block indicates the number of the partition between cpus.}\label{fig:hilbert-scf}
    \end{figure}

	An example is shown in figure \ref{fig:hilbert-scf} to demonstrate the procedure, in which the green dots represent the centre of the blocks to be partitioned and the line connecting them indicates the order. 

	\subsection{Communication between CPUs}
	In each CPU, an array of derived type object `CMT' in Fortran is defined to communicate with all of its neighbor CPUs. It contains the following data:
	\begin{itemize}
		\item An integer indicating the target CPU ID to communicate with;
		\item A pointer list pointting to the blocks of which the data should be packed and sent away;
		\item A pointer list pointting to the blocks of which the data should be received from the target process and be overwritten;
		\item Two buffer arrays to hold the data being sent and received.
	\end{itemize}
    For each communication object, there is a counterpart object in the corresponding neighboring CPU to form a communication pair.
	When implementting the communication, the `CMT' object firstly pack the data from the blocks to be sent into the send buffer. Then, the buffer is sent to the neighbor CPU with a non-blocking mpi\_send procedure. At the same time, another non-blocking mpi\_irecive procedure is called to recive the data from the counterpart `CMT'.
    These communication objects should be destroyed and recreated as long as the mesh is changed since a new partition of mesh is made. All the data needed to be sent away are packed into an array before the communication. When the recipient has the package received, the data are unpacked and stored in the corresponding blocks.
	\subsection{Basic procedure}
	Typically, the package can be used in different organizations for different purpose. However, some principles should be followed to establish a new solver framework.

\begin{figure}
		\centering
		\includegraphics*[width=\linewidth]{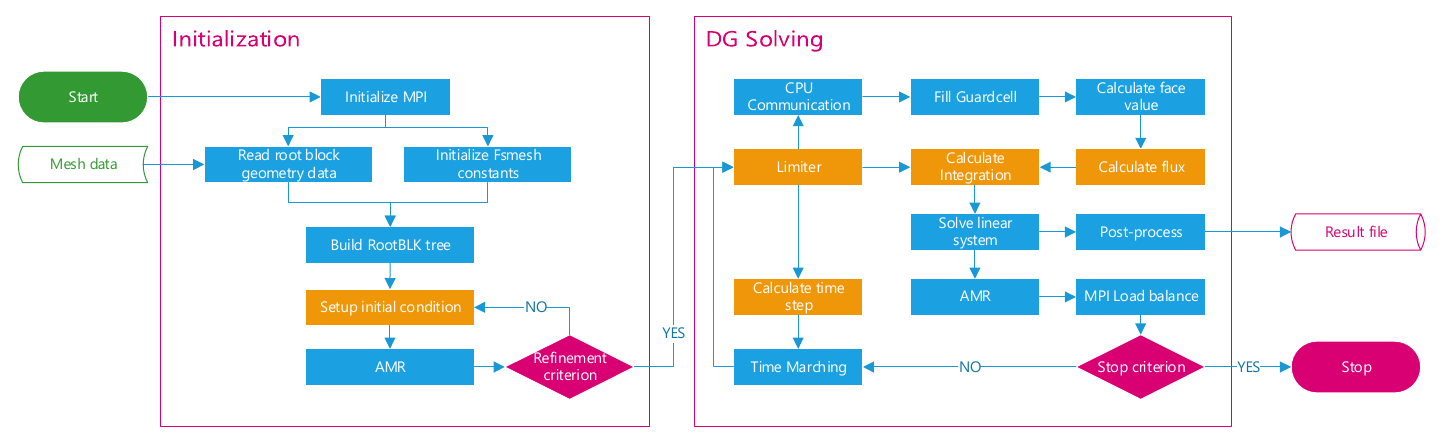}
		\caption{Sketch for the global procedure of a typical usage of the package. The yellow blocks denote the user-defined subroutines or functions.}
		\label{fig:workflow}
	\end{figure}

	As an example, here we provide a procedure for the ordinary RKDG for reference:
	\begin{enumerate}[label=(\arabic*)]
		\item Initialization
		\begin{enumerate}
			\item Initialize the MPI package;
			\item Initialize the global constants for the FSMesh package;
			\item Read from external files to initialize the root blocks;
			\item Iteratively initialize the unknowns with the given initial condition, calculate the refinement derefinement mark with the criterion, and do the refinement and balance the load in different threads.
		\end{enumerate}
		\item Solve the DG formulation for a single time increment
		\begin{enumerate}
			\item Evaluate the critical time increment among all the blocks with the CFL condition;
			\item Communicate between threads to fill the buffering blocks;
			\item Dump solution to the temporary array to ensure data consistency when filling guarding cells;
			\item Communicate between blocks to fill the guarding cells;
			\item Calculate cell face values and fluxes.
			\item Calculate flux and internal volume integrations to solve the DG formula for each block and update the solution;
		\end{enumerate}
		\item Check the refinement criterion and do the refinement.
		\item Go back to step 2 if the stop criterion is not met.
	\end{enumerate}
	The procedure is also shown in figure \ref{fig:workflow}, where only the task in yellow blocks requires user defined subroutines and the others will be completed by the package. 
	\section{Test cases}\label{sect:cases}
	In this section, we will show some test cases to demonstrate the ability of the present package. 
	\subsection{Linear convection equation}\label{sect:convection}
	Firstly, we consider the most simple case, i.e., a linear scalar convection equation reading
	\begin{equation}
		\frac{\partial u}{\partial t} + \mathbf{v}\cdot \nabla u =0.
	\end{equation}
	Here, $\mathbf{v}$ is the convection velocity which is taken as 1 in each direction for simplicity.
	We test the package in both 2- and 3-dimensional cases. The initial condition is given by
	\begin{equation}
			u(0,x,y) = [\cos(2\pi x)-1][\cos(2\pi y)-1]
	\end{equation}
	for the 2-dimensional case, while
	\begin{equation}
		u(0,x,y,z) = [\cos(4\pi x)-1][\cos(4\pi y)-1][\cos(4\pi z)-1]
	\end{equation}
	for the 3-dimensional case.
	The periodic boundary conditions are applied to the external boundaries.
	The computational domain is chosen as $[0,1]^{n_d}$. 

\begin{figure}
		\centering
		\includegraphics[width=0.95\linewidth]{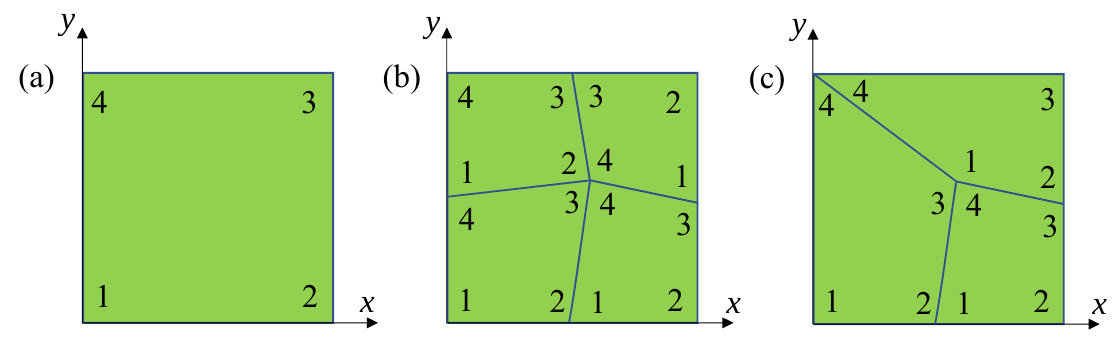}
		\caption{Sketch for the two sets of root blocks for the test case 1. (a) a single square root block; (b) 4 root blocks with shape distortion; (c) 3 root blocks with relative rotation.}\label{fig:test-convect-mesh}
	\end{figure}

	For the 2-dimensional case, three sets of root blocks are used to test the capability of the method. The first set only contains a single square root block covering the whole computational domain, the second set contains 4 root blocks with distorted block shape, and the third set contains 3 root blocks with unstructured connection, as shown in figure \ref{fig:test-convect-mesh}. The refinement criteria parameter is simply chosen as the average value of the DG cell, and the upper and lower criteria are set to 3.0 and 1.0, respectively. The third-order Runge-Kutta method is adopted as the time-marching scheme.

\begin{figure}
		\centering
		\subfigure[]{\includegraphics[width=0.32\linewidth]{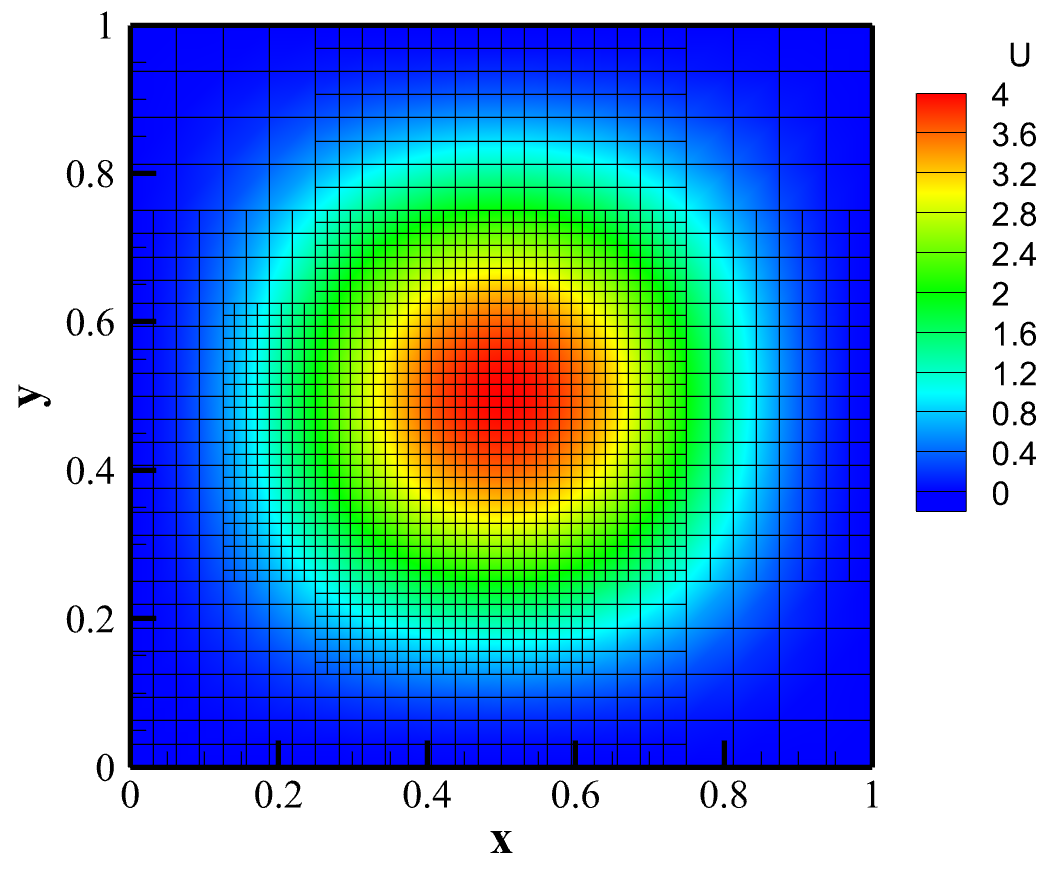}}
		\subfigure[]{\includegraphics[width=0.32\linewidth]{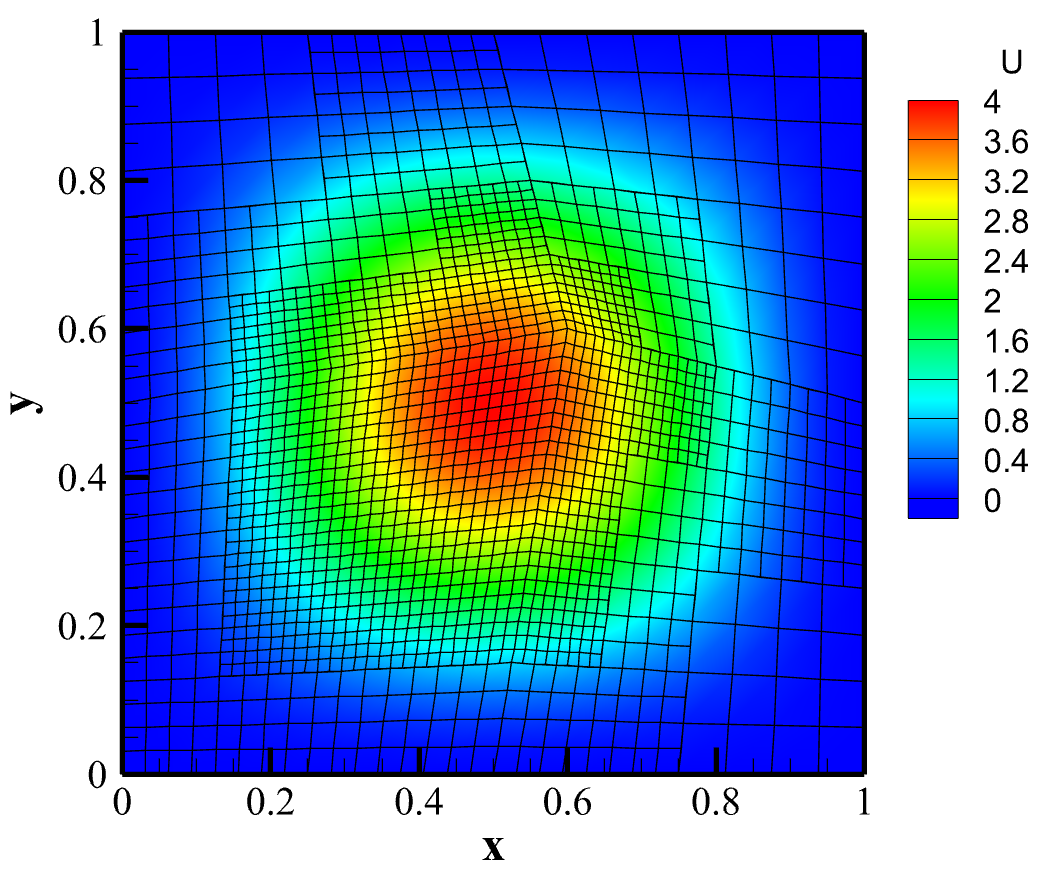}}
		\subfigure[]{\includegraphics[width=0.32\linewidth]{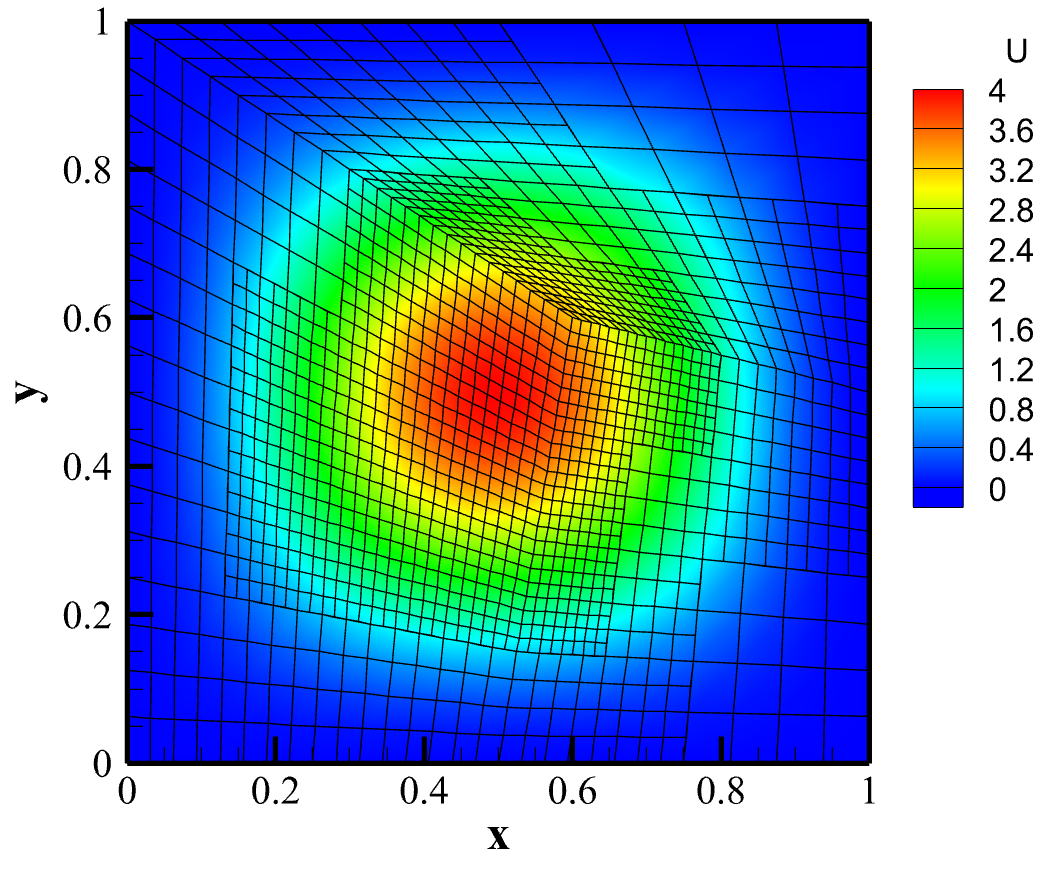}}
		\caption{Results of the 2-dimensional convection equation at $t = 1.0$  with different initial root block setup. (a) a single square root block; (b) 4 root blocks with shape distortion; (c) 3 root blocks with relative rotation.}\label{fig:test-convect-result}
	\end{figure}

	The results at $t = 1.0$ are given in figure \ref{fig:test-convect-result} with the maximum and minimum refinement levels set to 3 and 5 for the fist set grid, while 2 and 4 for the rest two sets such that the averaged cell sizes are similar. The black lines represent the cell boundaries. The results agree with the exact solution well with the $L_2$ errors being $2.2e^{-8}$, $2.7e^{-8}$ and $6.4e^{-8}$, respectively, which indicates the framework is capable of treating distored and unstructured mesh without losing accuracy.

\begin{figure}
		\centering
		\subfigure[]{\includegraphics[width=0.32\linewidth]{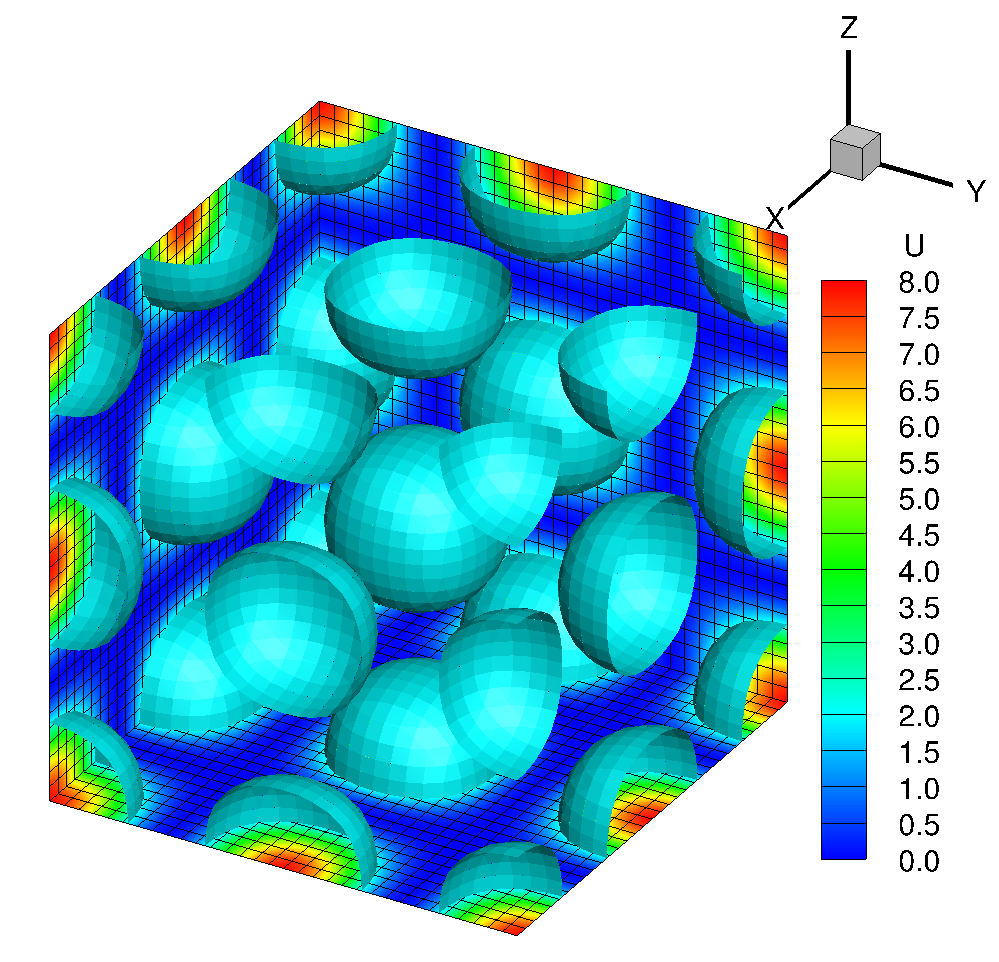}}
		\subfigure[]{\includegraphics[width=0.32\linewidth]{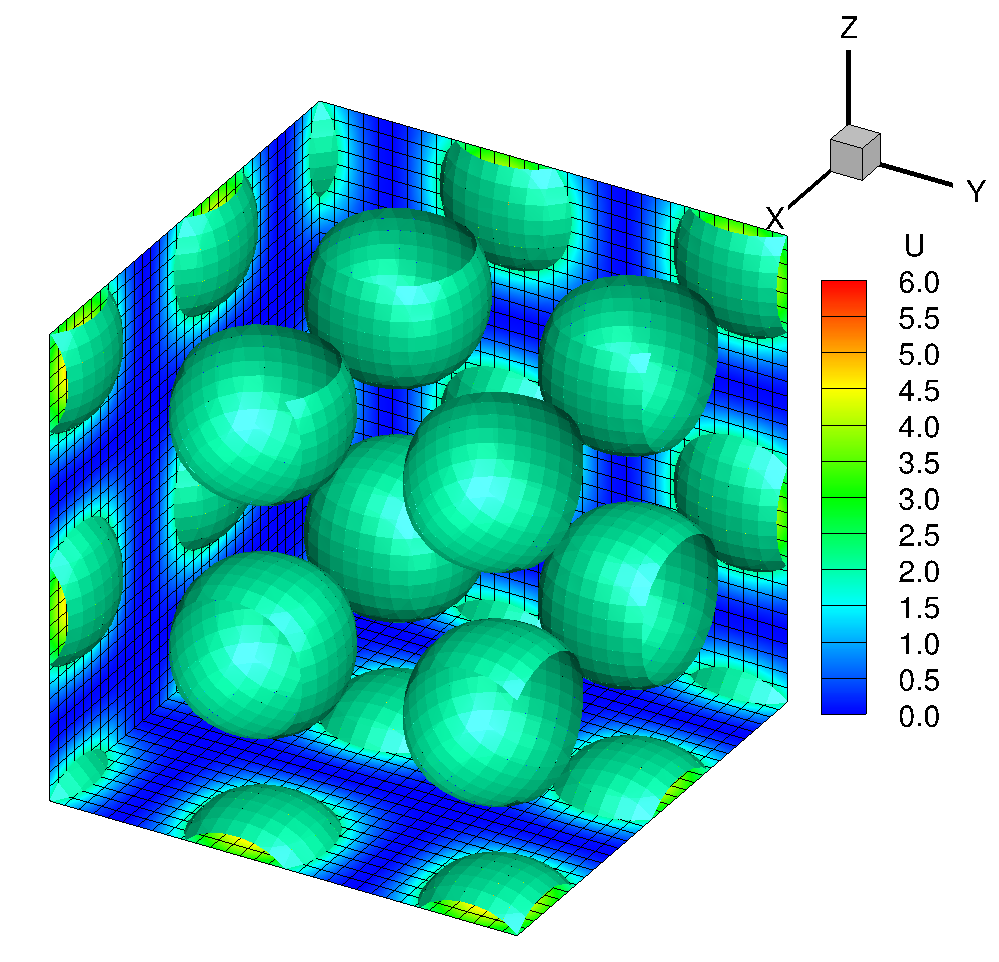}}
		\subfigure[]{\includegraphics[width=0.32\linewidth]{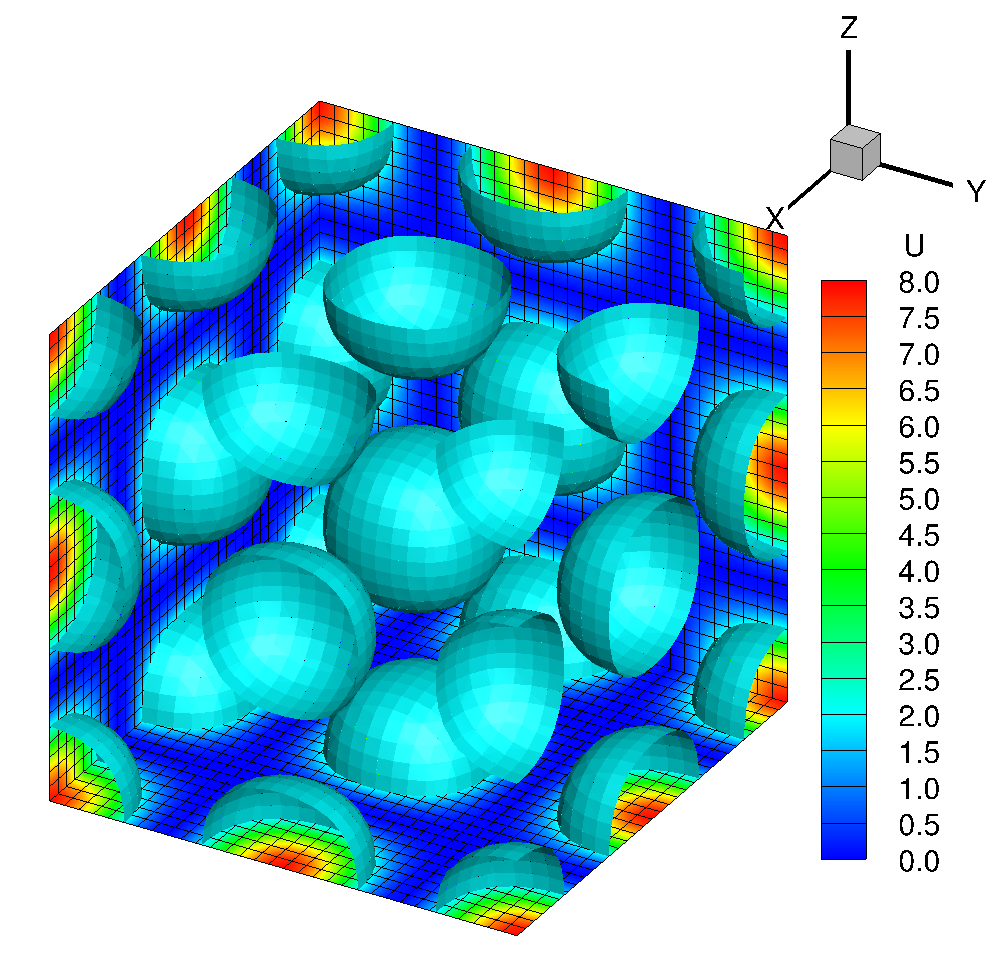}}
		\caption{Results of the 3-dimensional convection equation at $t = 0, 0.4$ and $0.5$, respectively. The solution slices at $x=0$, $y = 0$ and $z=0$ are shown in each subfigure. The iso-surface at $U = 2.0$ is also plotted.}
		\label{fig:test-convect-result-3d}
	\end{figure}

	For the 3-dimensional case, we cover the computitional domain with a single root block and refined it to 4 levels. The results at $t = 0.25$ is given in figure \ref{fig:test-convect-result-3d}, in which the result slices at $x=0$, $y=0$ and $z=0$ are shown in each subfigure. The iso-surface at $U = 2.0$ is also plotted. The results agree with the exact solution well, with the $L_2$ errors being $2.3e^{-4}$.

	\subsection{Diffusion equation}
    Now we consider solving the diffusion equation as
	\begin{equation}
		\frac{\partial u}{\partial t} - \nabla\cdot(\nabla u) = 0
	\end{equation}
	which is of great interest in various physical situations. The second-order derivative is discretized with the IPDG method\cite{SIAM_IPDG}. We solved the two-dimensional problem on the computational domain as $[0,1]\times [0,1]$ with Dirichlet boundary conditions. The boundary  and initial conditions are both set with
	\begin{equation}
		u(t,x,y) = e^{ax+by+(a^2+b^2)t}
	\end{equation}
	which satisfies the control equation analytically. 
    We used a single square root block covering the whole computational domain and took $a= -1$ and $b=1$, respectively. Then we choose $N_{seg}=6 $. The maximum and minimum refinement levels are set to 4 and 2, respectively. In contrast, the upper and lower average value refinement criteria are set to 3.0 and 1.0, respectively.
    
\begin{figure}[h]
       \centering
       \includegraphics[width=0.4\linewidth]{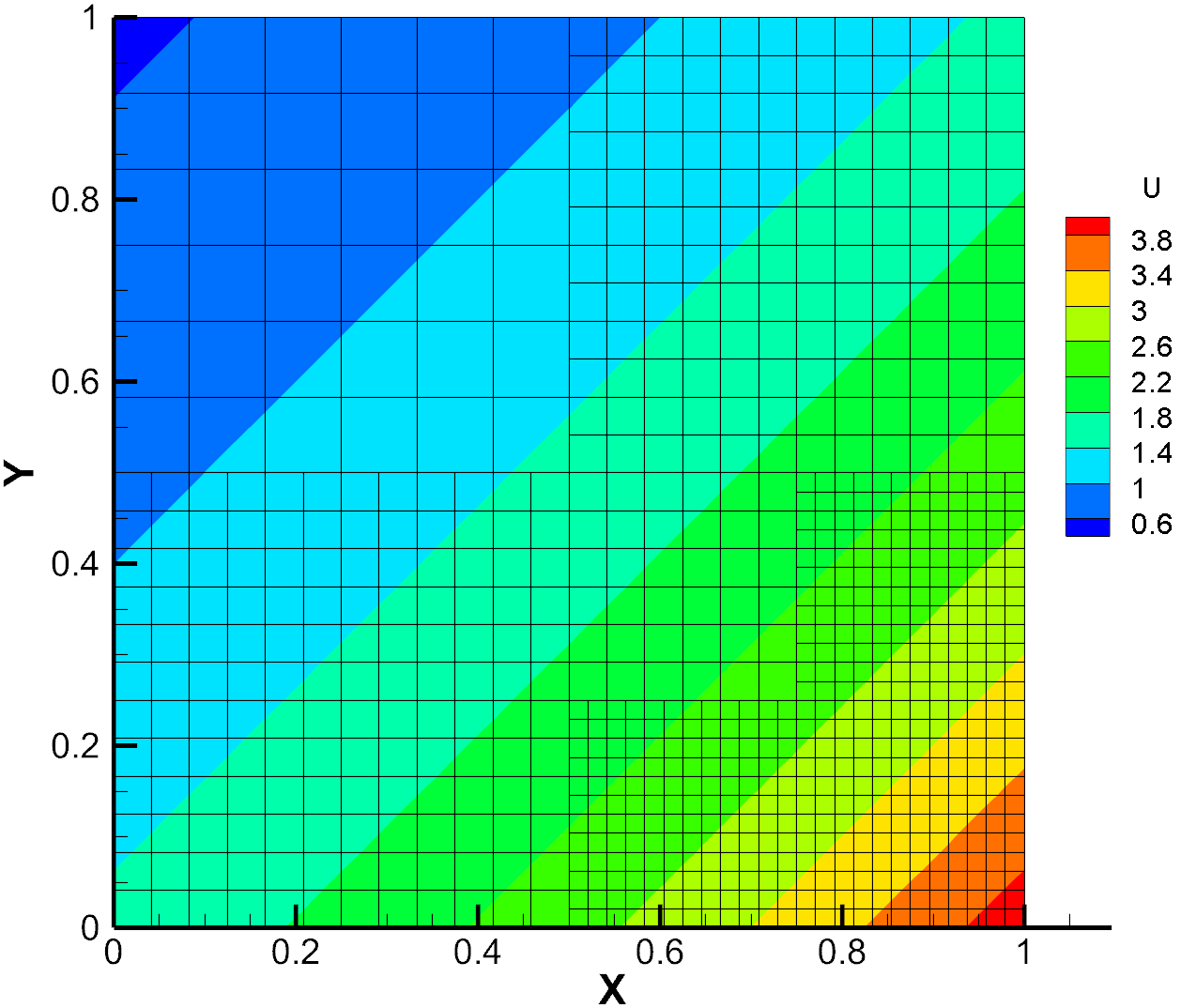}
       \caption{Result of the Diffusion problem at $t=0.2$ with the adaptive grids.}
       \label{fig:diffuse}
    \end{figure}

    The result at $t=0.2$ is given in figure \ref{fig:diffuse}. The result agrees with the exact solution well, with the $L_2$ error being  $2.3e^{-5}$. The above result shows that this framework is also feasible and efficient in dealing with diffusion equations.

	\subsection{Euler equation}
	
	\subsubsection{Cylindrical Sedov blast problem}
    	The first Euler equation test case is the two-dimensional Sedov blast problem. In this case, the explosion energy is suddenly released at a single point inside a still gaseous fluid flow, which is featured by low-density region and strong shocks. The analytic solution can be given by the self-similarity analysis\cite{Sedov_blast_Taylor}. To close the Euler equation, the ideal gas equation is adopted in this case which reads
		\begin{equation}
			p = \rho e (\gamma -1),
		\end{equation}
		where $\gamma$ is the material constant taken as 1.4 here.

	    The initial condition is given by
		\begin{equation}
			(\rho,u,v,E,\gamma)=\left\{
			\begin{array}{rcl}
				(1.0,0.0,0.0,10^{-12},1.4)\hfill& 							{|x|>\Delta x , |y|>\Delta y}\\
				(1.0,0.0,0.0,\frac{0.244816}{\Delta x\Delta y},1.4)\hfill&	\text{otherwise}
			\end{array} \right. 
		\end{equation}
		where ${\Delta x}$ and ${\Delta y}$ are the cell sizes in $x$ and $y$ directions, respectively. The computational domain is $[-1.1,1.1]\times [-1.1,1.1]$ and the final computation time is $t = 1.0$.

	 Only a single root block covers the whole computational domain, and the outflow boundary condition is applied to the external boundary. Then we set $N_{seg} = 10$ with the maximum and minimum refinement level are set to 4 and 6, respectively. Thus, the minimum size of the grid in the computational domain is 1.1/40, and the maximum size is 1.1/160. The upper and lower refinement criterias are set to 1.2 and 1.01, respectively. 

\begin{figure}
		\centering
		\subfigure[]{\includegraphics[height=0.4\linewidth]{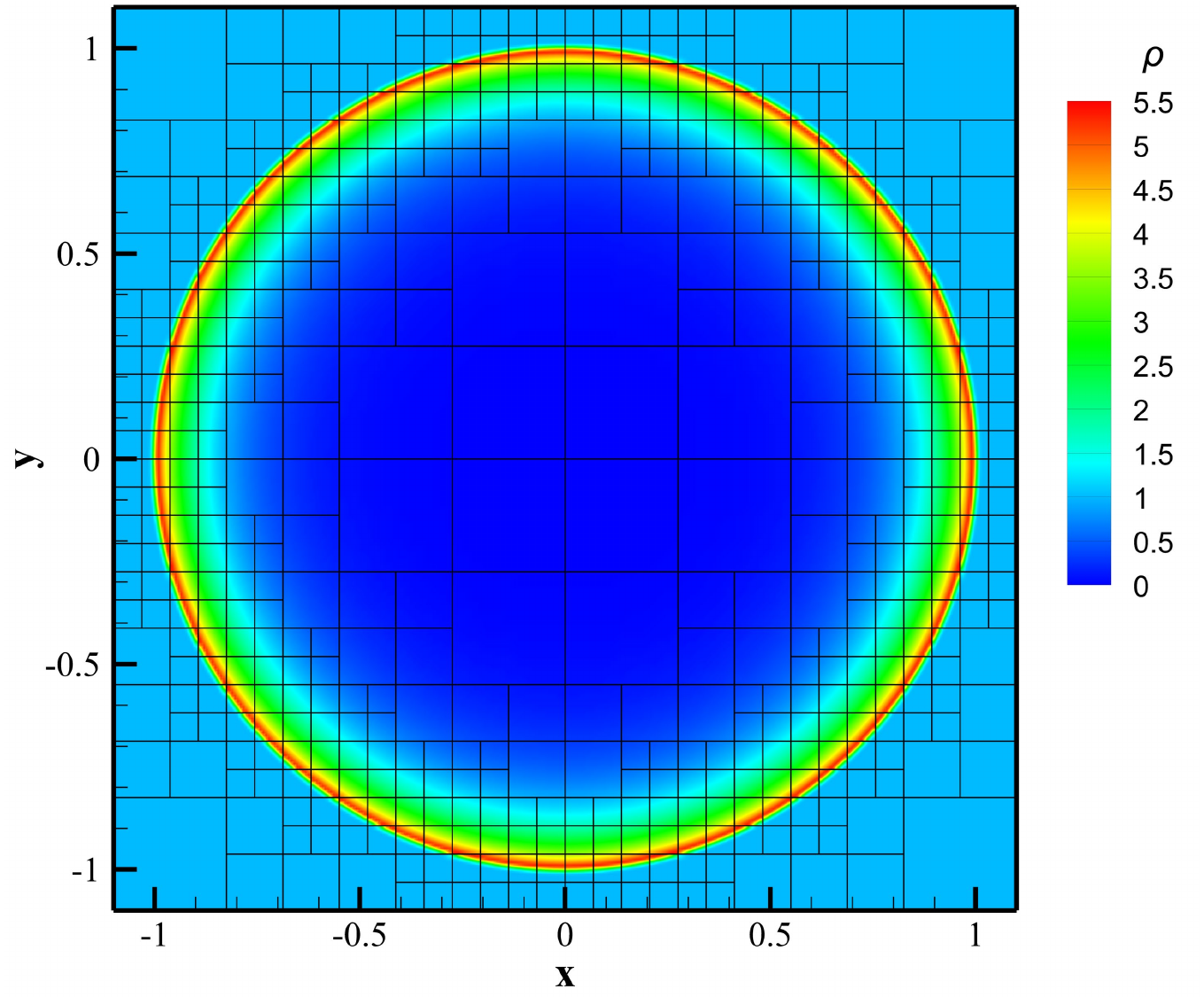}}
		\subfigure[]{\includegraphics[height=0.4\linewidth]{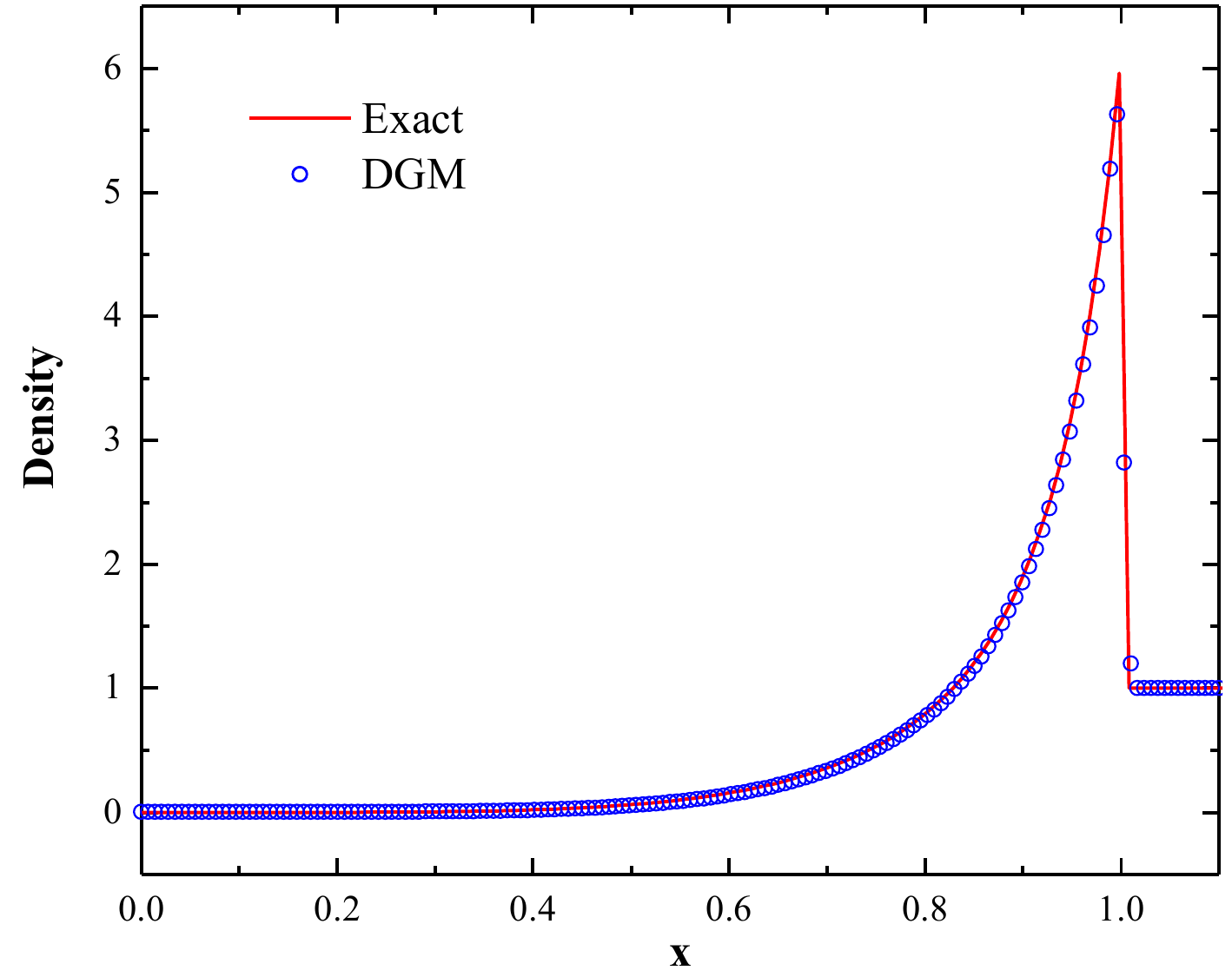}}
		\caption{Results of the Sedov balst problem at $t = 1.0$  with the adaptive grids. (a) density contour image with adaptive blocks; (b) density profile along $y = 0$.}\label{fig:test-sedov-result}
	\end{figure}

	The results at $t = 1.0$ for the Sedov blast wave are given in Figure \ref{fig:test-sedov-result}. The contour image and profile of density show that the numerical density agrees well with the exact solution, and the distribution of adaptive blocks also proves that the refined region can track the discontinuity of the shock wave to obtain high-resolution solution around it.

	\subsubsection{Flow around a blunt body}\label{sect:cylinder}
    The flow around a blunt body is the third Euler equation test case. The details of the flow field can be used to test the accuracy of the numerical method and the validity of the adaptive mesh refinement framework proposed in this paper. 
    We consider a cylinder of radius $R=0.2$ with its center at $(0,0)$ in a supersonic flow field, where the initial condition of the flow field and the inflow boundary condition are given by

	\begin{equation}
		(\rho,u,v,p,\gamma)=(1.4,\text{Ma},0.0,1.0,1.4),
	\end{equation}
in which Ma is Mach number, which is taken as $1.8$ in this paper.

The square computational domain is $[-1.0,2.6]\times [-1.8,1.8]$ in size and consists of $320$ initial root blocks, with supersonic gas flowing from left to right. Except that the left side of the computational domain is the inflow boundary, the rest are the outflow boundary. Moreover, the range of the refinement levels is set to $1 \sim 5$ with $N_{seg} = 4$. The boundary of the cylinder is refined to the finnest level, and the node correction strategy is used to maintain smoothness in refining.

\begin{figure}
		\centering
		\subfigure[]{\includegraphics[height=0.4\linewidth]{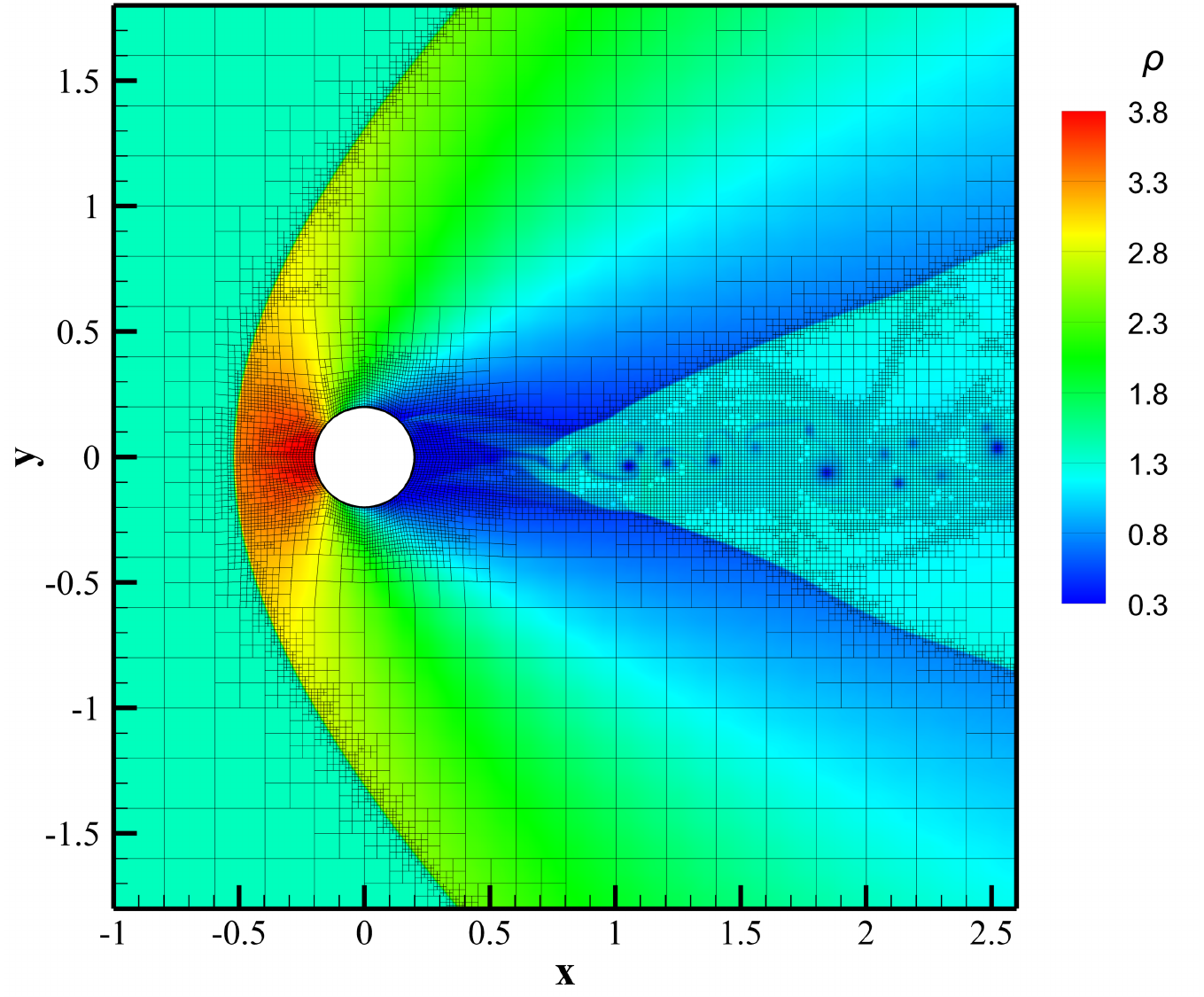}}\hspace{5mm}
		\subfigure[]{\includegraphics[height=0.4\linewidth]{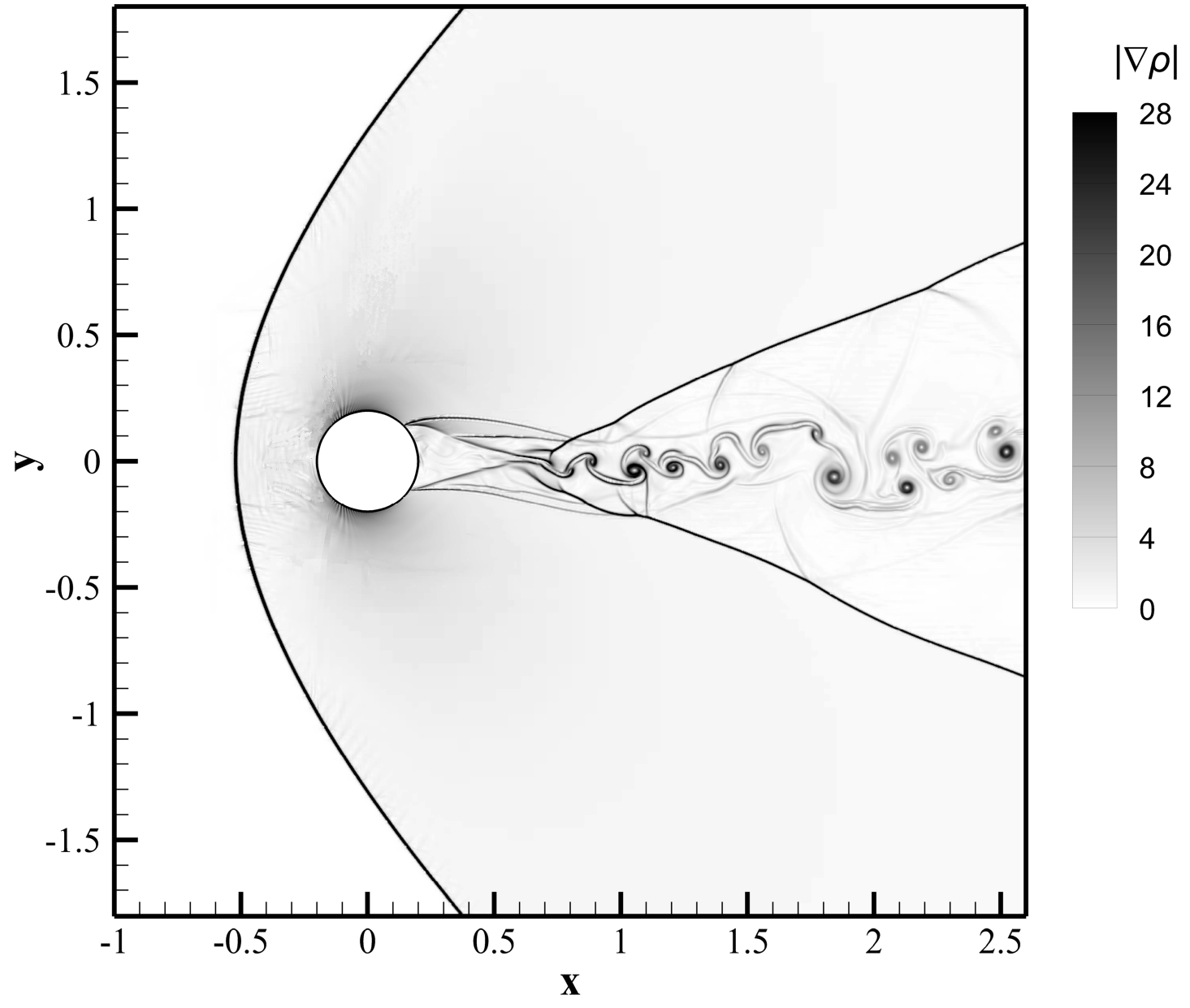}}
		\caption{(a)Density and (b)numerical schlieren diagram results of a supersonic flow around a cylinder with Ma = 1.8.}\label{fig:test-cylinder-result}
	\end{figure}

In this simulation, the refinement criteria is taken as the velocity divengence $|\frac{\partial u}{\partial x}+\frac{\partial v}{\partial y}|$ and the velocity curl $|\frac{\partial v}{\partial x}-\frac{\partial u}{\partial y}|$ to track shock waves and vortex structure in the flow field respectively. Furthermore, the positivity-preserving\cite{JCP_PP_limiter_2010,JCP_PP_limiter_2011} and WENO limiters\cite{JCP_WENO_limiter_2013} are adopted to ensure the parabolicity of the system and suppress the numerical oscillation.
From Figure \ref{fig:test-cylinder-result}, it can be seen that the bow detached shock wave, separation shock wave, and vortex generated by the shock wave interactions have been finely simulated, and the refinement region can be well attached to the above structures under our framework.

	\subsection{Efficiency test}
	In order to test the MPI parallel efficiency of this framework, a linear convection equation in Section \ref{sect:convection} is selected for parallel computation test, and the number of central processing units is set to $N_{\mathrm{CPUs}}=2^k$ with $k$ ranging from $0$ to $9$.
	The whole computational domain is covered by a single square root block as shown in Figure \ref{fig:test-convect-mesh}(a), which is refined seven times, resulting in a total of $16384$ blocks and $262144$ cells with $N_{seg} = 4$. And the calculation stops when $t=0.5$.

\begin{figure}[ht]
		\centering
		\includegraphics[height=0.4\linewidth]{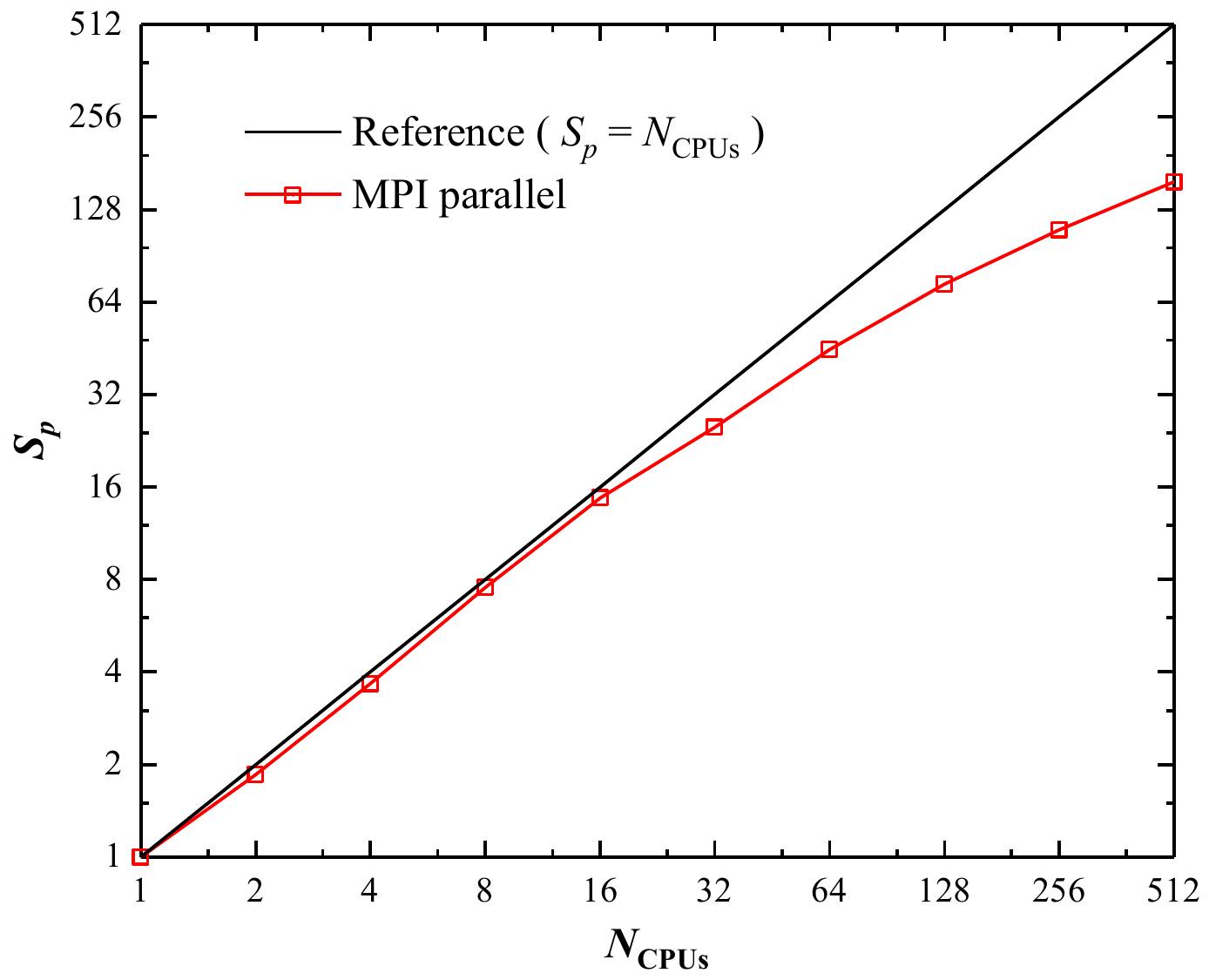}
		\caption{The logarithmic relationship of the acceleration ratio to the number of central processing units.}\label{fig:test-CPUs}
	\end{figure}

	The computing time $T_{N_{\mathrm{CPUs}}}$ and other data obtained are shown in Table \ref{tab:computing-time}, where $S_p$ represents the speedup of current $N_{\mathrm{CPUs}}$ with respect to $N_{\mathrm{CPUs}}=1$, defined as $S_p=\frac{T_{1}}{ T_{\mathrm{CPUs}}}$. It can be seen that a larger $N_{\mathrm{CPUs}}$ can greatly save computing costs. However, more time cost is required for data communication between central processing units as the $N_{\mathrm{CPUs}}$ increases, so the acceleration effect does not remain linear, and the effect of parallel acceleration gradually decreases, as shown in Figure \ref{fig:test-CPUs}. In general, the AMR framework in this paper has a good extensibility in MPI parallel computing, and can greatly improve the computing efficiency.

	\begin{table}[ht]
		\caption{Parallel computing time under different $N_{\mathrm{CPUs}}$.}\label{tab:computing-time}
		\centering
            \small
            \setlength{\tabcolsep}{20pt}
		\begin{tabular}{cccc}
			\toprule[1.0pt]
			$ N_{\mathrm{CPUs}} $ & $T_{N_{\mathrm{CPUs}}}/\rm s$ & $S_p$ & ${\rm log_2} (S_p)$\\
			\hline
			$1$    & 7034    & 1.00     & 0.00\\
			$2$    & 3797    & 1.85     & 0.89\\
                $4$    & 1919    & 3.67     & 1.87\\
                $8$    & 932     & 7.55     & 2.92\\
                $16$   & 478     & 14.72    & 3.88\\
                $32$   & 281     & 25.03    & 4.65\\
                $64$   & 157     & 44.80    & 5.49\\
                $128$  & 96      & 73.27    & 6.20\\
                $256$  & 64      & 109.91   & 6.78\\
                $512$  & 44.62  & 157.64    & 7.30\\
			\bottomrule[1.0pt]
		\end{tabular}
	\end{table}

	\section{Conclusion}\label{sect:conclusion}
	A new adaptive mesh refinement approach designed for the discontinuous Galerkin method is presented. With the block being the basic element of AMR, each block can be recrusively split into self-similar children blocks with the same data structure, which forms a quad- or oct-tree. The tree structures grown from different root blocks connected unstructuredly to generate a so called forest according to proper refinement and derefinement cretirions. In such a way, the ideas of the localized structured grid inside each block and the unstructured topology between root blocks are combined. Thus, the data structure is significantly simplified while maintaining the ability to deal with complex geometries. 

	The communication between neighbor blocks are implemented by filling the guard cells surrounding the blocks. The guard cells are placed around the blocks to provide external solution data to calculate the face fluxes at the block boundary. The $L_2$ projection of the solution from the real cells of the source block into the guard cells of the destination block is used to ensure the global conserveness and high-order property. In such a way, the direct calculation of numerical fluxes at non-conformal faces is avoided and only a structured DG grid inside each block is needed to be solved. 
	Besides, the smooth external boundaries of the computational domain with and without an explicit surface function can both be treated to avoid angular boundaries while refining.

	The present AMR method has been implemented in the FSMesh Package and been applied into several 2- and 3-dimensioanl benchmark cases. The results suggested that the optimal converging rate can be archived in smooth problems and the shock wave front and vortexes can be easily tracked and refined dynamically to obtain high-resolution results.

    The package presented in this paper is available on reasonable request by contacting yunlong\_liu@hrbeu.edu.cn.

	\section*{Acknowledgments}
	This work was supported by the National Natural Science Foundation of China (Grant No. xx).

\bibliography{reference}
\end{document}